\documentclass[aps,twocolumn ]{revtex4}
\usepackage{amsfonts}
\usepackage{amsmath}
\usepackage{amssymb,epsf}
\usepackage{graphicx,color}

\begin{document}

\title{Dilatonic black holes in gravity's rainbow with a nonlinear source:\\
the effects of thermal fluctuations}
\author{S. H. Hendi$^{1,2}$\footnote{
email address: hendi@shirazu.ac.ir}, B. Eslam
Panah$^{1,2,3}$\footnote{
email address: behzad.slampanah@gmail.com}, S. Panahiyan$^{1,4,5}$ \footnote{%
email address: shahram.panahiyan@uni-jena.de} and M. Momennia$^{1}$ \footnote{%
email address: m.momennia@shirazu.ac.ir}} \affiliation{$^1$
Physics Department and Biruni Observatory, College of Sciences,
Shiraz
University, Shiraz 71454, Iran\\
$^2$ Research Institute for Astronomy and Astrophysics of Maragha (RIAAM),
P.O. Box 55134-441, Maragha, Iran\\
$^3$ICRANet, Piazza della Repubblica 10, I-65122 Pescara, Italy\\
$^4$Helmholtz-Institut Jena, Fr\"{o}belstieg 3, D-07743 Jena, Germany\\
$^5$Physics Department, Shahid Beheshti University, Tehran 19839, Iran}

\begin{abstract}
This paper is devoted to investigate nonlinearly charged dilatonic black
holes in the context of gravity's rainbow with two cases: I) by considering
the usual entropy, II) in the presence of first order logarithmic correction
of entropy. First, exact black hole solutions of dilatonic Born{Infeld
gravity with an energy dependent Liouville-type potential are obtained.
Then, thermodynamic properties of the mentioned cases are studied,
separately. It will be shown that although mass, entropy and heat capacity
are modified due to the presence of first order correction, the temperature
remains independent of it. Furthermore, it will be shown that divergencies
of the heat capacity, hence phase transition points are also independent of
first order correction whereas the stability conditions are highly sensitive
to variation of the correction parameter. Except for the effects of first
order correction, we will also draw a limit on values of the dilatonic
parameter and show that it is possible to recognize AdS and dS
thermodynamical behaviors for two specific branches of the dilatonic
parameter. In addition, the effects of nonlinear electromagnetic field and
energy functions on thermodynamical behavior of the solutions will be
highlighted and dependency of critical behavior, on these generalizations
will be investigated. }
\end{abstract}

\maketitle

\section{Introduction}

Albeit extremely successful consequences of Einstein's theory of gravity,
there are various reasons which signal that this theory should be modified,
at least at high energy regime. Regarding the physical processes at energies
of order of the Planck scale, one is expected to modify Einstein gravity to
a quantum theory of gravity which is related to Quantum Field Theory and
could be the correct framework to describe processes at ultraviolet (UV)
energies. Considering the UV-completion regime of black holes, the
gravitational interaction at energies exceeding the Planck mass implies that
the UV-completion must be achieved by new quantum degrees of freedom of
wavelength much shorter than the Planck length. However, it is expected to
recover the semi-classical (thermodynamical) behavior of black holes in the
mean-field approximation \cite{Mean1,Mean2}. Horava-Lifshitz gravity \cite%
{Horava} is one of the interesting UV modifications of general relativity
(GR), in which reduces to GR at infrared (IR) limit. Based on Lifshitz
scaling for space and time, different attractive subjects have been
considered such as string theory of types IIA and IIB \cite{Gregory,Burda},
AdS/CFT correspondence \cite{Gubser,Ong,Alishahiha}, dilatonic black
holes/branes \cite{Tarrio,Lee,Greenwald,Goldstein,Bertoldi} and phase
transitions/geometrical thermodynamics of black holes \cite%
{Cao,biswas,Quevedo}.

On the other hand, another point of view regarding the UV completion of GR
is related to gravity's rainbow \cite{MagueijoI,MagueijoII}. In order to
introduce the gravity's rainbow, we can start from its historical origin;
the doubly special relativity. It is known that the special theory of
relativity is based on two postulates \cite{Einstein}; (i) the laws of
physics are invariant in all inertial frames, and (ii) the constant
invariant speed of light in the vacuum for all observers, regardless of the
relative motion of the light source. This results into the fact that the
speed of a massive particle cannot be equal or larger than the speed of
light. Based on high energy point of view, if one consider an upper bound
for the energy of a test particle and add this assumption as a postulate to
the others mentioned above, the doubly special relativity will be
constructed \cite{MagueijoA,MagueijoB,MagueijoC}. So, in this manner, the
energy of test particle cannot be greater than the Planck energy. Following
the gravity supplementing to the special relativity which leads to GR, one
can generalize doubly special relativity in the presence of gravity to
obtain the so-called doubly general relativity or gravity's rainbow \cite%
{MagueijoII}.

Another way for constructing such theories is considering a modification of
the standard energy-momentum relation. The modification of usual
energy-momentum relation in gravity's rainbow is given as
\begin{equation}
E^{2}f^{2}\left( \varepsilon \right) -p^{2}g^{2}\left( \varepsilon \right)
=m^{2},  \notag
\end{equation}%
where $E$ and $E_{p}$ are the energy of a test particle and the Planck
energy, respectively and they are related to together with $\varepsilon
=E/E_{p} $ in which $\varepsilon \leq 1$, because the energy of a test
particle $E$ cannot be greater than $E_{p}$ \cite{Peng}. The functions $%
f\left( \varepsilon \right) $ and $g\left( \varepsilon \right) $ are called
rainbow functions. They are phenomenologically motivated and could be
extracted by experimental data \cite{AliK2015}, see Refs. \cite%
{MagueijoI,Jacob,AmelinoLiv} for more details. It is worthwhile to mention
that the rainbow functions are required to satisfy $\underset{\varepsilon
\rightarrow 0}{\lim }f\left( \varepsilon \right) =1$ and $\underset{%
\varepsilon \rightarrow 0}{\lim }g\left( \varepsilon \right) =1$, where
these conditions ensure that the modified energy-momentum relation reduces
to its usual form in the IR limit. It is worth mentioning that such
justification is based on the standard model of particle physics. Since
quantum theories of gravity are different from classical one in high energy
regime, a full quantum theory has to be valid in UV regime. It is notable
that both of the Horava-Lifshitz gravity and gravity's rainbow are valid in
the UV regime. On the other hand, considering suitable choice of the energy
functions, the Horava-Lifshitz gravity can be related to the gravity's
rainbow \cite{Garattini}. For the reasons mentioned, we consider the
gravity's rainbow in this paper.

Recently, there has been a growing interest in gravity's rainbow because of
its interesting achievements in the context of theoretical physics, such as
providing possible solutions for information paradox \cite{Ali,Gim},
admitting usual uncertainty principle \cite{Ling,Li}, existence of remnants
for black holes after evaporation \cite{AliFaizal,AliPRD}, and absence of
black hole production at LHC \cite{absence}. In addition, from cosmological
point of view, it is possible to remove the big bang singularity by using
gravity's rainbow \cite{Awad,Santos,HendiNon,Momennia}. Black hole
thermodynamics in the presence of gravity's rainbow coupled to (non)linear
electrodynamics has been discussed in literature \cite%
{HendiENrainbow,Banerjee,Feng}. Moreover, the effects of rainbow functions
on the thermodynamics properties and phase transition of black holes have
been studied in Gauss-Bonnet gravity \cite{HendiGBRain}, Kaluza-Klein theory
\cite{Alsaleh}, massive gravity \cite{HendiMassrain}, and $F(R)$ theories of
gravity \cite{GarattiniII,HendiFRrain}. In the view point of astrophysics,
the structure of neutron stars have been investigated and it is found that
by increasing the rainbow function more than one, the maximum mass of
neutron stars increases \cite{HendiNeutronRain1,HendiNeutronRain2}.

Regarding various aspects of black holes in the context of gravity's
rainbow, it will be interesting to look for possible coupling of rainbow
functions with scalar field and its self interacting potential. Astronomical
observations indicated that our Universe has an accelerated expansion \cite%
{PerlmutterI,Riess}. Einstein gravity cannot explain such acceleration. In
this regards, various modifications of Einstein gravity are proposed in
literature. Among them, scalar tensor gravity and string theory are most
acceptable ones with interesting properties. It is worth mentioning that the
low energy limit of effective string theory could lead to appearance of a
dilaton scalar field \cite{Witten,HorneH} which motivates one to investigate
GR coupled with such a scalar field. In addition, dilaton field coupled with
other gauge fields has significant effects on the solutions \cite%
{Koikawa,Gibbons,Brill,Garfinkle}. In Ref. \cite{Choi}, it was shown that
dilaton field can be an appropriate candidate for dark matter. In addition,
in order to have deep picture regarding the nature of dark energy, a new
scalar field can be added to the field content of original theory \cite%
{HuangI,HuangII}. Also, GR coupled with a dilaton field can explain the
accelerated expansion of our Universe, properly. Moreover, it was shown that
in the presence of dilaton scalar field, the asymptotic behavior of black
hole spacetimes is neither flat nor (a)dS \cite%
{Mignemi,ChanHorne,CaiDilaton,ClementPRD,SheykhiPLB}. This is due to the
fact that in the limit $r\rightarrow \infty $, the effects of dilaton field
is still present. Besides, it was shown that the dilatonic black hole
solutions can be constructed in the (a)dS spacetime background \cite{Gao}.
Black objects in the presence of dilaton gravity and their thermodynamics
have been studied in Refs. \cite%
{DBHII,DBHIII,DBHIV,DBHV,DBHVI,DBHVII,DBHVIII,DBHX,DBHXII,DBHXIII}.
Recently, considering the dilaton field, the hydrostatic equilibrium
equation of compact objects has been obtained and the properties of neutron
stars have been analyzed in Refs. \cite{Fiziev,HendiNeutron}. The
evaporation of quantum black holes has also been investigated using two
dimensional dilaton gravity \cite{CallanG,Banks}. In addition, there is a
strong motivation to study rainbow deformation of geometries that occur in
the string theory.

Nonlinear field theory is one of the most interesting branches in physical
sciences because most physical systems presented in the nature are
nonlinear. On the other hand, existence of some limitations in the linear
Maxwell theory motivates one to consider nonlinear electrodynamics (NED)
\cite{Heisenberg,Schwinger,Lorenci,Novello}. One of the main advantages of
considering NED theories comes from the fact that these theories are richer
than the linear Maxwell theory and in some special cases they reduce to the
Maxwell field. Besides, it was shown that NED can remove the black hole and
big bang singularities \cite{BeatoGarcia,LorenciNovello,CordaMosquera}. In
addition, the effects of NED become quite important in superstrongly
magnetized compact objects \cite{MosqueraSalim,Birula}. Studying general
relativity coupled to NED is an attractive subject because of its specific
properties in gauge/gravity duality. One of the most important NED theories
is so-called Born-Infeld NED (BI NED) which has been introduced by Born and
Infeld in $1934$ \cite{Born}. This interesting type of NED removes the
divergence point of electric field of a point-like charge and such a
property was the main motivation for introducing this kind of NED. Other
strong motivation for considering BI NED comes from the fact that it
naturally arises in the low-energy limit of open string theory \cite%
{Fradkin,Bergshoeff,Callan,Leigh}. The following incomplete list shows the
investigation of gravitational fields coupled to BI NED for static black
holes \cite%
{Dehghani,Zou,Mazharimousavi,Myung,Fernando,CaiPang,Wirschins,Cataldo},
wormholes \cite{Lu,DehHendi}, rotating black objects \cite{Ferrari,HendiBS},
and superconductors \cite{Jing}.

Motivated by these applications, in this paper, we analyze dilaton field
using the formalism of gravity's rainbow in the presence of BI NED. So in
the present paper, we are going to study the dilatonic black hole solutions
in Einstein gravity's rainbow coupled to BI NED with considering the effects
of thermal fluctuations.

\section{Black hole solutions of dilaton gravity's rainbow in the presence
of BI NED}

Here, we are interested in dilatonic charged black holes in the context of
gravity's rainbow with BI source. Since we are interested in minimal
coupling, the action can be written as \cite{Chan}
\begin{equation}
I=\frac{1}{16\pi }\int d^{4}x\sqrt{-g}\left[ R-2\left( \nabla \Phi \right)
^{2}-V\left( \Phi \right) -L\left( F,\Phi \right) \right] ,  \label{Action}
\end{equation}%
where $R$ is the Ricci scalar, and $\Phi $ and $V(\Phi)$ are, respectively,
dilaton field and its corresponding potential. $L( F,\Phi)$ is the
Lagrangian of the BI electromagnetic field under consideration with the
following explicit form
\begin{equation}
L\left( F,\Phi \right) =4\beta ^{2}e^{2\alpha \Phi }\left( 1-\sqrt{1+\frac{%
e^{-4\alpha \Phi }F}{2\beta ^{2}}}\right) ,
\end{equation}%
where the Maxwell invariant is denoted by $F=F_{\mu \nu }F^{\mu \nu }$ ($%
F_{\mu \nu }=\partial _{\mu }A_{\nu }-\partial _{\nu }A_{\mu }$ in which $%
A_{\mu }$ is the gauge potential). Also, $\beta $ is the nonlinearity
parameter and dilatonic constant is identified by $\alpha$ which is a
parameter for determining the strength of coupling between the scalar field
and electrodynamics.

For the sake of simplification in calculation, we use the following
redefinition
\begin{equation}
L\left( F,\Phi \right) =4\beta ^{2}e^{2\alpha \Phi }L\left( Y\right) ,
\end{equation}
where $L(Y)=1-\sqrt{1+Y}$ and $Y=\frac{e^{-4\alpha \Phi }F}{2\beta ^{2}}$.

Variation of the action (\ref{Action}) leads to the following field
equations
\begin{eqnarray}
R_{\mu \nu }&=&2\left( \partial _{\mu }\Phi \partial _{\nu }\Phi +\frac{%
g_{\mu \nu }}{4}V(\Phi )\right) -4e^{-2\alpha \Phi }\partial _{Y}L\left(
Y\right) F_{\mu \eta }F_{\nu }^{\eta }  \notag \\
&+& 2\beta ^{2}e^{2\alpha \Phi }\left[ 2Y\partial _{Y}L\left( Y\right)
-L\left( Y\right) \right] g_{\mu \nu },  \label{dilaton
equation(I)}
\end{eqnarray}
\begin{equation}
\nabla ^{2}\Phi =\frac{1}{4}\frac{\partial V}{\partial \Phi }+2\alpha \beta
^{2}e^{2\alpha \Phi }\left[ 2Y\partial _{Y}L\left( Y\right) -L\left(
Y\right) \right] ,  \label{dilaton equation(II)}
\end{equation}%
\begin{equation}
\nabla _{\mu }\left( e^{-2\alpha \Phi }\partial _{Y}L\left( Y\right) F^{\mu
\nu }\right) =0.  \label{Maxwell equation}
\end{equation}

Here, we consider the following energy dependent line element
\begin{equation}
ds^{2}=-\frac{\Psi (r)}{f^{2}(\varepsilon )}dt^{2}+\frac{1}{%
g^{2}(\varepsilon )}\left[ \frac{dr^{2}}{\Psi (r)}+r^{2}R^{2}(r)d\Omega
_{k}^{2}\right] ,  \label{metric}
\end{equation}%
in which $R(r)$ and $\Psi(r)$ are metric functions which should be
determined. We recall that these two functions $g(\varepsilon )$ and $%
f(\varepsilon )$ are energy (or rainbow) functions that are chosen
phenomenologically. In addition, $d\Omega _{k}^{2}$ is the metric of two
dimensional subspace which depends on topology of the boundary of spacetime
and could have positive (elliptic $k=1$), zero (flat $k=0$) or negative
(hyperbolic $k=-1$) curvature such as
\begin{equation}
d\Omega _{k}^{2}=\left\{
\begin{array}{cc}
d\theta ^{2}+\sin ^{2}\theta d\varphi ^{2}, & k=1 \\
d\theta ^{2}+\sinh ^{2}\theta d\varphi ^{2}, & k=-1 \\
d\theta ^{2}+d\varphi ^{2}, & k=0%
\end{array}%
\right. .  \label{dOmega}
\end{equation}

Since we are interested in black holes with radial electric field, the
suitable choice of gauge potential is
\begin{equation}
A_{\mu }=\delta _{\mu }^{t}h\left( r\right) ,  \label{electric po}
\end{equation}%
which by using Eq. (\ref{Maxwell equation}), we can obtain the following
field equation
\begin{eqnarray}
&&\beta ^{2}e^{4\alpha \Phi }\left[ R(r)\left( r\alpha F_{tr}\Phi ^{\prime }-%
\frac{r}{2}F_{tr}^{\prime }-F_{tr}\right) -rF_{tr}R^{\prime }(r)\right]
\notag \\
&&+ f^{2}(\varepsilon )g^{2}(\varepsilon )F_{tr}^{3}\left[ R(r)+rR^{\prime
}(r)\right] =0,  \label{AuEq}
\end{eqnarray}
where $F_{tr}=A_{t}^{\prime }$ is the $tr$-component of the electromagnetic
tensor field and prime denotes $d/dr$. Solving Eq. (\ref{AuEq}), we obtain
\begin{equation}
F_{tr}=\frac{qe^{2\alpha \Phi }}{r^{2}R^{2}(r)\sqrt{1+\frac{%
q^{2}f^{2}(\varepsilon )g^{2}(\varepsilon )}{\beta ^{2}r^{4}R^{4}(r)}}},
\end{equation}
where $q$ is an integration constant which is related to the total electric
charge of the solutions. In order to find the metric functions, we consider
the following modified version of the Liouville-type dilaton potential
\begin{equation}
V(\Phi )=\frac{2k\alpha ^{2}}{b^{2}\Theta _{-1,1}}g^{2}(\varepsilon )e^{%
\frac{2\Phi }{\alpha }}+2\Lambda e^{2\alpha \Phi },  \label{V(Phi)}
\end{equation}%
where $\Lambda $ is a free parameter which plays the role of cosmological
constant. It is worthwhile to mention that in IR limit ($g(\varepsilon
)=f(\varepsilon )=1$), Eq. (\ref{V(Phi)}) reduces to the known
Liouville-type dilaton potential that is employed for finding
Friedman-Robertson-Walker scalar field cosmologies \cite{Ozer} and
Einstein-Maxwell-dilaton black holes \cite{Chan,YazadjievCQG}.

Here, we consider $R(r)=e^{\alpha \Phi (r)}$ as an ansatz for finding the
metric function. This ansatz is supported by studies conducted in
Einstein-Maxwell-dilaton gravity \cite{DehghaniF}.

Now, by using Eqs. (\ref{dilaton equation(I)}), (\ref{dilaton equation(II)})
and the metric (\ref{metric}) with obtained electromagnetic field tensor,
one can find the following differential equation for calculating the metric
function, analytically,
\begin{eqnarray}
&&\frac{kr^{2}\alpha ^{2}}{b^{2}\Theta _{-1,1}}\left( \frac{b}{r}\right) ^{%
\frac{\gamma }{\alpha ^{2}}}+\frac{r\Theta _{1,1}\Psi ^{\prime }(r)-\Theta
_{-1,1}\Psi (r)}{\Theta _{1,1}^{2}}-k\left( \frac{b}{r}\right) ^{\gamma }
\notag \\
&&+\frac{2r^{2}}{g^{2}(\varepsilon )}\left( \frac{b}{r}\right) ^{3\gamma }%
\left[ \beta ^{2}\left( 1+\sqrt{1+\eta }\right) +\frac{\Lambda }{2}\right] =0
\end{eqnarray}%
where $\Theta _{i,j}=i+j\alpha ^{2}$, $\eta =\frac{q^{2}f^{2}(\varepsilon
)g^{2}(\varepsilon )}{\beta ^{2}r^{4}}\left( \frac{r}{b}\right) ^{2\gamma }$
and $\gamma =2\alpha ^{2}/\Theta _{1,1}$. Using obtained field equations,
one can find both the metric function and dilatonic field in following forms
\begin{eqnarray}
\Psi (r) &=&-\frac{\Theta _{1,1}}{\Theta _{-1,1}}k\left( \frac{b}{r}\right)
^{-\gamma }-\frac{m}{r^{1-\gamma }}+\frac{\Lambda \Theta _{1,1}^{2}b^{\gamma
}}{g^{2}(\varepsilon )\Theta _{-3,1}}r^{(2-\gamma )}  \notag \\
&&-\frac{2\beta ^{2}\Theta _{1,1}^{2}b^{\gamma }r^{(2-\gamma )}}{%
g^{2}(\varepsilon )\Theta _{-3,1}}\left[ 1-H_{1}\right] ,  \label{Psi} \\
&&  \notag \\
\Phi (r) &=&\frac{\alpha }{\Theta _{1,1}}\ln \left( \frac{b}{r}\right) ,
\label{Phi(r)}
\end{eqnarray}%
\newline
where $H_{1}={}_{2}F_{1}\left( \left[ -\frac{1}{2},\frac{\Theta _{-3,1}}{4}%
\right] ,\left[ \frac{\Theta _{1,1}}{4}\right] ,-\eta \right) $ is the
hypergeometric function. Also, $b$ is an arbitrary constant related to the
scalar field and $m$ is an integration constant which is related to total
mass of the solutions. It is worthwhile to mention that for limiting case of
$\beta \longrightarrow \infty $, obtained metric function will lead to
\begin{equation}
\Psi (r)=-\frac{k\Theta _{1,1}}{\Theta _{-1,1}}\left( \frac{r}{b}\right)
^{\gamma }-\frac{m}{r^{1-\gamma }}+\frac{\Lambda \Theta _{1,1}^{2}b^{\gamma
}r^{(2-\gamma )}}{g^{2}(\varepsilon )\Theta _{-3,1}}+\frac{%
q^{2}f(\varepsilon )^{2}\Theta _{1,1}}{r^{2}}(\frac{r}{b})^{\gamma },
\end{equation}%
which is charged black hole in gravity's rainbow \cite{HendiFEP}. On the
other hand, for limiting case of $\alpha =\gamma =0$ and $\beta
\longrightarrow \infty $, our solutions reduce to
\begin{equation}
\Psi (r)=k-\frac{m}{r}-\frac{\Lambda }{3}\frac{r^{2}}{g^{2}(\varepsilon )}+%
\frac{q^{2}f^{2}(\varepsilon )}{r^{2}},
\end{equation}%
which is the metric function of $4$-dimensional asymptotically AdS
topological charged black hole in gravity's rainbow \cite{HendiFRrain}.

In order to study the effects of matter fields as well as dilatonic gravity,
we will investigate the behavior of the Kretschmann scalar for small and
large values of radial coordinate. The existence of divergency for this
scalar denotes that our solution contains a curvature singularity. If this
singularity is covered with a horizon, obtained solution is interpreted as
black hole. It is a matter of calculation to show that for this black hole,
we have
\begin{eqnarray}
\lim_{r\rightarrow 0}R_{\alpha \beta \mu \nu }R^{\alpha \beta \mu \nu }
&\propto &r^{-\frac{4\Theta _{2,1}}{\Theta _{1,1}}},  \label{RR0} \\
\lim_{r\rightarrow \infty }R_{\alpha \beta \mu \nu }R^{\alpha \beta \mu \nu
} &=&\frac{12\Lambda^2 (\alpha ^{4}-2\alpha ^{2}+2)}{\Theta _{3,-1}^{2}}%
\left( \frac{b}{r}\right) ^{4\gamma }.  \label{RRinf}
\end{eqnarray}

\begin{figure}[tbp]
$%
\begin{array}{c}
\epsfxsize=7cm \epsffile{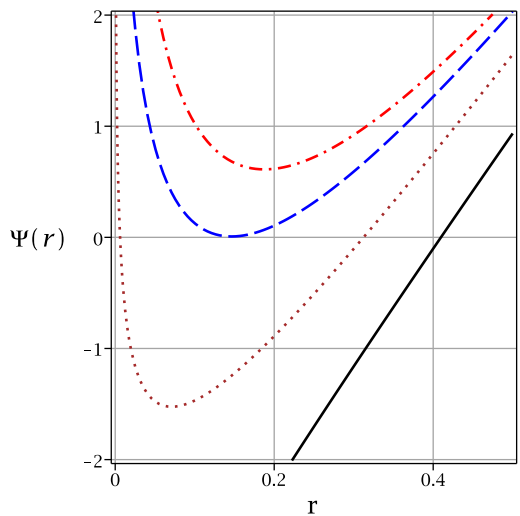}%
\end{array}
$%
\caption{$\Psi(r)$ versus $r$ for $k=1$, $m=5$, $\Lambda=-1$, $b=1.2$ and $%
q=1$.\newline
$\protect\alpha=0.9$, $f(\protect\varepsilon)=g(\protect\varepsilon)=0.9$, $%
\protect\beta=0.1$ (continues line), $\protect\beta=0.3$ (dotted line), $%
\protect\beta=0.534$ (dashed line) and $\protect\beta=0.7$ (dashed-dotted
line).}
\label{Fig1}
\end{figure}

It is evident from Eq. (\ref{RR0}) that there is an essential singularity
located at the origin for this solution. Therefore, the first condition for
having black hole is satisfied. On the other hand, the asymptotical behavior
of system is modified due to dilatonic gravity and it is not (A)dS. It is
notable that, the existence of horizon for the solution is investigated
through following diagrams (Fig. \ref{Fig1}).

Plotted diagram shows that depending on the choices of different parameters,
obtained solutions may present black holes with two horizons, extremal black
holes and naked singularity. For exampe, in the case of BI theory, for small
values of nonlinearity parameter observed behavior is Schwarzschild like
(continues line of Fig. \ref{Fig1}). Increasing nonlinearity parameter will
change the behavior of the metric function into Rissner-Nordstrom like which
may yield two horizons (dotted line of Fig. \ref{Fig1}), extremal black
holes (dashed line of Fig. \ref{Fig1})\ or naked singularity (dashed-dotted
line of Fig. \ref{Fig1}). Here, we saw that generalization to nonlinear
electrodynamics as well as gravity's rainbow, provided a richer
phenomenologies for black holes. We see that geometrical structure of the
black holes have been modified due to these generalizations. Depending on
the choices of different parameters, the type of singularity, the general
behavior of metric function and number and type of horizons have been
modified. These modifications highlight the contributions of this matter
field and gravities. We will continue to study the effects of these
generalizations on conserved and thermodynamical quantities in the next
section.

\section{Thermodynamics}

\subsection{Usual thermodynamics (non-correction)}

This section is devoted to the calculation of conserved quantities without
thermal fluctuations. Here, we investigate the effects of gravity's rainbow,
dilaton scalar field and nonlinear BI electrodynamics on thermodynamical
quantities and check the validity of the first law of black hole
thermodynamics. Using the concept of surface gravity, one can show that the
Hawking temperature for obtained solutions will be
\begin{widetext}
\begin{equation}
T=\frac{\Theta _{1,1}\left( \frac{b}{r_{+}}\right) ^{\gamma
}\left[ \Theta
_{1,-1}r_{+}^{2}\left( \beta ^{2}-\frac{\Lambda }{2}\right) -\frac{%
kg^{2}(\varepsilon )}{2}\left( \frac{b}{r_{+}}\right) ^{-2\gamma }-\Theta
_{1,-1}\beta ^{2}r_{+}^{2}\sqrt{1+\eta _{+}}\right] }{2\pi \Theta
_{1,-1}r_{+}f(\varepsilon )g(\varepsilon )},  \label{temp}
\end{equation}
\end{widetext}
where $\eta _{+}=\eta _{\left\vert _{r=r_{+}}\right. }$. Since we are
working in Einstein framework, the entropy of these black holes could be
obtained by the area law
\begin{equation}
S=\frac{A}{4}=\frac{r_{+}^{2}\left( \frac{b}{r_{+}}\right) ^{\gamma }}{%
4g^{2}(\varepsilon )},  \label{entropy}
\end{equation}%
in which by setting $\alpha =0$ and $g(\varepsilon )=1$, the entropy of
black holes in Einstein gravity is recovered. In order to find the total
electric charge of these black holes, one can use the Gauss's law which
leads to following result
\begin{equation}
Q=\frac{qf(\varepsilon )}{4\pi g(\varepsilon )}.  \label{Q}
\end{equation}

The electric potential of the black holes at the horizon radius with respect
to spacial infinity as a reference could be calculated with the following
relation
\begin{equation}
U\left( r\right) =\left. A_{\mu }\chi ^{\mu }\right\vert _{r\longrightarrow
\infty }-\left. A_{\mu }\chi ^{\mu }\right\vert _{r\longrightarrow r_{+}},
\label{defU}
\end{equation}
where for nonlinear BI theory can be obtained as
\begin{equation}
U=\frac{q}{r} \;{}_{2}F _{1}\left( \left[ -\frac{1}{2},\frac{\Theta _{1,1}}{4%
}\right] ,\left[ \frac{\Theta _{5,1}}{4}\right] ,-\eta _{+}\right) .
\label{U}
\end{equation}

In order to obtain the total mass of black holes, one can use the definition
of Abbott and Deser \cite{AD1,AD2,AD3}, which leads to
\begin{equation}
M=\frac{b^{\gamma }}{8\pi \Theta _{1,1}f(\varepsilon )g(\varepsilon )}m.
\label{Mass}
\end{equation}

It is worthwhile to mention that for IR limit ($g(\varepsilon
)=f(\varepsilon )=1$) and by setting $\alpha =0$, Eq. (\ref{Mass}) reduces
to the mass of Einstein-Maxwell black holes.

Now, we are in a position to examine the validity of the first law of
thermodynamics. To do so, first, we should obtain geometrical mass of the
solutions, $m$, as a function of other parameters, by using $\Psi\left(
r=r_{+}\right) =0$. Then, by replacing it in Eq. (\ref{Mass}), we obtain a
relation for the total mass of the black holes versus entropy and electric
charge. In this case, the extensive parameters will be the entropy and total
electric charge and their corresponding conjugating quantities are the
temperature and the electric potential, respectively. Therefore, the
validation of the first law of black holes thermodynamics is done by
\begin{equation}
\left( \frac{\partial M}{\partial S}\right) _{Q}=T\text{ \ \ \ \ }\&\text{\
\ \ \ \ }\left( \frac{\partial M}{\partial Q}\right) _{S}=U.  \label{AS}
\end{equation}

It is a matter of the calculation to show that obtained quantities in Eq. (%
\ref{AS}) are exactly the same with those obtained in Eqs. (\ref{temp}) and (%
\ref{U}). Therefore, we find that the first law is valid as
\begin{equation}
dM=\left( \frac{\partial M}{\partial S}\right) _{Q}dS+\left( \frac{\partial M%
}{\partial Q}\right) _{S}dQ.
\end{equation}

Our final subject of the interest in this section is heat capacity. The
information that are provided by this quantity could be used to render the
thermodynamical structure of black holes in the context of their thermal
stability/instability. In addition, the existence of discontinuity for this
quantity signals the presence of thermal phase transition. It is worthwhile
to mention that the discontinuities in the heat capacity usually are
observed. For black holes in canonical ensemble, the heat capacity is
calculated by following relation
\begin{equation}
C=T\left( \frac{\partial S}{\partial T}\right) _{Q},  \label{heat}
\end{equation}%
in which by using obtained temperature (\ref{temp}) and entropy (\ref%
{entropy}), one can find it for this specific thermodynamical quantities as
\begin{widetext}
\begin{equation}
C=\frac{r_{+}^{4}\left( \frac{b}{r_{+}}\right) ^{\gamma }\left[ \frac{%
kg^{2}(\varepsilon )}{2\Theta _{-1,1}}\left( \frac{b}{r_{+}}\right)
^{-2\gamma }+\beta ^{2}r_{+}^{2}\eta _{+}-r_{+}^{2}\left( \beta ^{2}-\frac{%
\Lambda }{2}\right) \right] \eta _{+}}{\Theta _{-1,1}g^{2}(\varepsilon )%
\left[ r_{+}^{2}\left( \frac{kg^{2}(\varepsilon )}{\Theta _{-1,1}\left(
\frac{b}{r_{+}}\right) ^{2\gamma }}+2r_{+}^{2}\left( \beta ^{2}-\frac{%
\Lambda }{2}\right) \right) \eta _{+}-2\beta ^{2}r_{+}^{4}-\frac{2\Theta
_{1,1}q^{2}f^{2}(\varepsilon )g^{2}(\varepsilon )}{\Theta _{-1,1}\left(
\frac{b}{r_{+}}\right) ^{2\gamma }}\right] }.  \label{heatNon}
\end{equation}
\end{widetext}
The thermodynamical stability and possible phase transition of the black
holes in this case will be discussed in following sections. In the next
section, we will obtain the thermodynamical quantities for the case where
entropy is corrected to include a logarithmic correction.

\subsection{Thermal fluctuations: correction}

Here, we want to investigate the effects of thermal fluctuations on obtained
solutions. Consideration of thermal fluctuations results into modifications
of different thermodynamical quantities, though some of the quantities
remain fixed. To the leading order, the entropy gets logarithmic correction
and given by
\begin{equation}
S=S_{0}-\frac{\zeta }{2}\log \left( S_{0}T^{2}\right) ,  \label{correctEN}
\end{equation}%
in which $S_{0}$ is the uncorrected entropy which is given in the equation (%
\ref{entropy}). Also, $\zeta $ is the thermal fluctuation parameter which we
will call it correction parameter through the paper. Using Eqs. (\ref{temp})
and (\ref{entropy}), we can obtain the corrected entropy as
\begin{equation}
S=\frac{r_{+}^{2}\left( \frac{b}{r_{+}}\right) ^{\gamma }}{%
4g^{2}(\varepsilon)}-\frac{\zeta }{2}\log \left( \frac{r_{+}^{2}\left( \frac{%
b}{r_{+}}\right) ^{\gamma }}{4g^{2}(\varepsilon )}T^{2}\right) .
\label{Correctent}
\end{equation}

Our investigation regarding surface gravity confirms that the temperature is
one of the thermodynamical quantities which is not affected by the presence
of first order correction. In other words, for this case, the obtained
temperature will be same as that was previously obtained (\ref{temp}). Using
the corrected entropy (\ref{Correctent}) and temperature (\ref{temp}), we
are able to compute the Helmholtz free energy as
\begin{equation}
F=-\int SdT=\frac{1}{16\pi f(\varepsilon )g^{3}(\varepsilon )}\int \frac{\Pi
_{1}\Pi _{2}dr_{+}}{r_{+}^{2}\sqrt{1+\eta _{+}}},
\end{equation}
where
\begin{widetext}
\begin{equation}
\Pi _{1}=1-\frac{2g^{2}(\varepsilon )\zeta }{r_{+}^{2}\left( \frac{b}{r_{+}}%
\right) ^{\gamma }}\ln \left( \left[ \frac{\left( \frac{b}{r_{+}}\right) ^{%
\frac{3\gamma }{2}}\Theta _{1,1}}{4\pi f(\varepsilon )g^{2}(\varepsilon )}%
\left( \beta ^{2}r_{+}^{2}\sqrt{1+\eta _{+}}+\frac{kg^{2}(\varepsilon )}{%
2\Theta _{-1,1}}\left( \frac{b}{r_{+}}\right) ^{-2\gamma }-\frac{r_{+}^{2}}{2%
}\left( 2\beta ^{2}-\Lambda \right) \right) \right] ^{2}\right),
\nonumber
\end{equation}
\begin{equation}
\Pi _{2}=r_{+}^{2}\sqrt{1+\eta _{+}}\left[ kg^{2}(\varepsilon
)+r_{+}^{2}\left( \frac{b}{r_{+}}\right) ^{2\gamma }\Theta _{-1,1}\left(
2\beta ^{2}-\Lambda \right) \right] -2\left[ \Theta _{1,1}f^{2}(\varepsilon
)g^{2}(\varepsilon )q^{2}+\Theta _{-1,1}\beta ^{2}r_{+}^{4}\left( \frac{b}{%
r_{+}}\right) ^{2\gamma }\right].  \nonumber
\end{equation}
\end{widetext}
Now, by using the obtained Helmholtz free energy and employing its relation
with mass, $M=F+TS$, one can calculate the corrected mass in the following
form
\begin{widetext}
\begin{equation}
M=F-\frac{r_{+}\Pi _{1}\Theta _{1,1}}{16\pi f(\varepsilon )g^{3}(\varepsilon
)\Theta _{-1,1}}\left[ kg^{2}(\varepsilon )+2\beta ^{2}r_{+}^{2}\Theta
_{-1,1}\left( \frac{b}{r_{+}}\right) ^{2\gamma }\sqrt{1+\eta _{+}}%
-r_{+}^{2}\Theta _{-1,1}\left( \frac{b}{r_{+}}\right) ^{2\gamma }\left(
2\beta ^{2}-\Lambda \right) \right] .
\end{equation}
\end{widetext}

We obtain the electric charge ($Q$) and the electric (chemical) potential ($U
$), which are similar to Eqs. (\ref{Q}) and (\ref{U}), respectively.
Therefore, by using the obtained thermodynamic quantities, the first law of
black hole thermodynamics satisfy as
\begin{equation}
dM=TdS+UdQ.
\end{equation}

Finally, the heat capacity of this case is of interest. It is a matter of
calculation to obtain this quantity by using Eqs. (\ref{temp}) and (\ref%
{Correctent}) with (\ref{heat}) as
\begin{widetext}
\begin{eqnarray}
C &=&T\left( \frac{dS}{dT}\right) _{Q}=\frac{r_{+}^{4}\left( \frac{b}{r_{+}}%
\right) ^{\gamma }\left[ \frac{kg^{2}(\varepsilon )}{2\Theta _{-1,1}}\left(
\frac{b}{r_{+}}\right) ^{-2\gamma }+\beta ^{2}r_{+}^{2}\eta
_{+}-r_{+}^{2}\left( \beta ^{2}-\frac{\Lambda }{2}\right) \right] \eta _{+}}{%
\Theta _{-1,1}g^{2}(\varepsilon )\left[ r_{+}^{2}\left( \frac{%
kg^{2}(\varepsilon )}{\Theta _{-1,1}\left( \frac{b}{r_{+}}\right) ^{2\gamma }%
}+2r_{+}^{2}\left( \beta ^{2}-\frac{\Lambda }{2}\right) \right) \eta
_{+}-2\beta ^{2}r_{+}^{4}-\frac{2\Theta _{1,1}q^{2}f^{2}(\varepsilon
)g^{2}(\varepsilon )}{\Theta _{-1,1}\left( \frac{b}{r_{+}}\right) ^{2\gamma }%
}\right] }  \nonumber \\
&&  \nonumber \\
&&-\frac{\zeta \left[ r_{+}^{2}\left( \frac{k\alpha ^{2}g^{2}(\varepsilon
)\left( \frac{b}{r_{+}}\right) ^{-2\gamma }}{\Theta _{-1,1}}+2\Theta
_{-2,1}r_{+}^{2}\left( \beta ^{2}-\frac{\Lambda }{2}\right) \right) \eta
_{+}-2\left( \Theta _{-2,1}\beta ^{2}r_{+}^{4}+\frac{\alpha
^{2}q^{2}f^{2}(\varepsilon )g^{2}(\varepsilon )}{\left( \frac{b}{r_{+}}%
\right) ^{2\gamma }}\right) \right] }{\Theta _{-1,1}g^{2}(\varepsilon )\left[
r_{+}^{2}\left( \frac{kg^{2}(\varepsilon )}{\Theta _{-1,1}\left( \frac{b}{%
r_{+}}\right) ^{2\gamma }}+2r_{+}^{2}\left( \beta ^{2}-\frac{\Lambda }{2}%
\right) \right) \eta _{+}-2\beta ^{2}r_{+}^{4}-\frac{2\Theta
_{1,1}q^{2}f^{2}(\varepsilon )g^{2}(\varepsilon )}{\Theta _{-1,1}\left(
\frac{b}{r_{+}}\right) ^{2\gamma }}\right] }.  \label{heatCor}
\end{eqnarray}
\end{widetext}

Let us highlight some important properties of the obtained heat capacity in
the presence of first order correction. First of all, the effects of first
order correction are observed only in the numerator of heat capacity which
indicates that divergencies of the heat capacity, hence phase transition
points are independent of first order correction. On the other hand, since
the first order correction is present in numerator, it is expected that
roots of the heat capacity and stability conditions are first order
correction dependent. The contributions of first order correction and
differences between correction and non-correction cases will be discussed in
details in next section.

\subsection{Thermal structure: comparison between correction and
non-correction}

In this section, the main goal is to understand the possible scenarios
regarding thermal structure of the black hole solutions. Since we have
obtained thermodynamical properties in the context of both first order
correction and non-corrected solutions, we will give details in the context
of both of them. Furthermore, we also investigate the details of
contribution of the first order correction.

The obtained temperatures for correction and non-correction cases are the
same. On the other hand, from the first law of thermodynamics, we can see
that temperature is calculated as a function of fluctuation in internal
energy with changes of entropy. This shows that change in entropy due to
variation of the correction parameter results into modifications in internal
energy on a level that the temperature of this system remains fixed. In
other words, modifications of entropy is realized by internal energy in a
manner that temperature remains fixed. This provides us with a tool to
increase/decrease the entropy and internal energy without any concern
regarding the possible changes in temperature. This is in a manner,
isothermal like behavior, although further tests are required to establish
the isothermal nature of this property.

Returning to the obtained temperature, one can recognizes specific
properties for it: I) obtained temperature provides us with imposing a limit
on values that dilatonic parameter could acquire ($\alpha \neq 1$). Such
limitation prevents temperature to have divergent value, II) using this
limit, one can see that essentially two branches exist for dilatonic
parameter, hence temperature: $\alpha >1$ and $\alpha <1$. This is due to
fact that the signature of different terms in temperature would be opposite
for these two cases. Only exceptions are the cosmological constant and
purely nonlinear terms. Later, we will show how these two branches give
different pictures regarding the thermodynamics of these black holes, III)
dilatonic parameter is coupled with all presented terms in the temperature.
This shows that dilatonic gravity has profound effects on the behavior of
temperature, hence thermodynamical structure of the black holes.

\begin{figure*}[tbp]
$%
\begin{array}{ccc}
\epsfxsize=5.5cm \epsffile{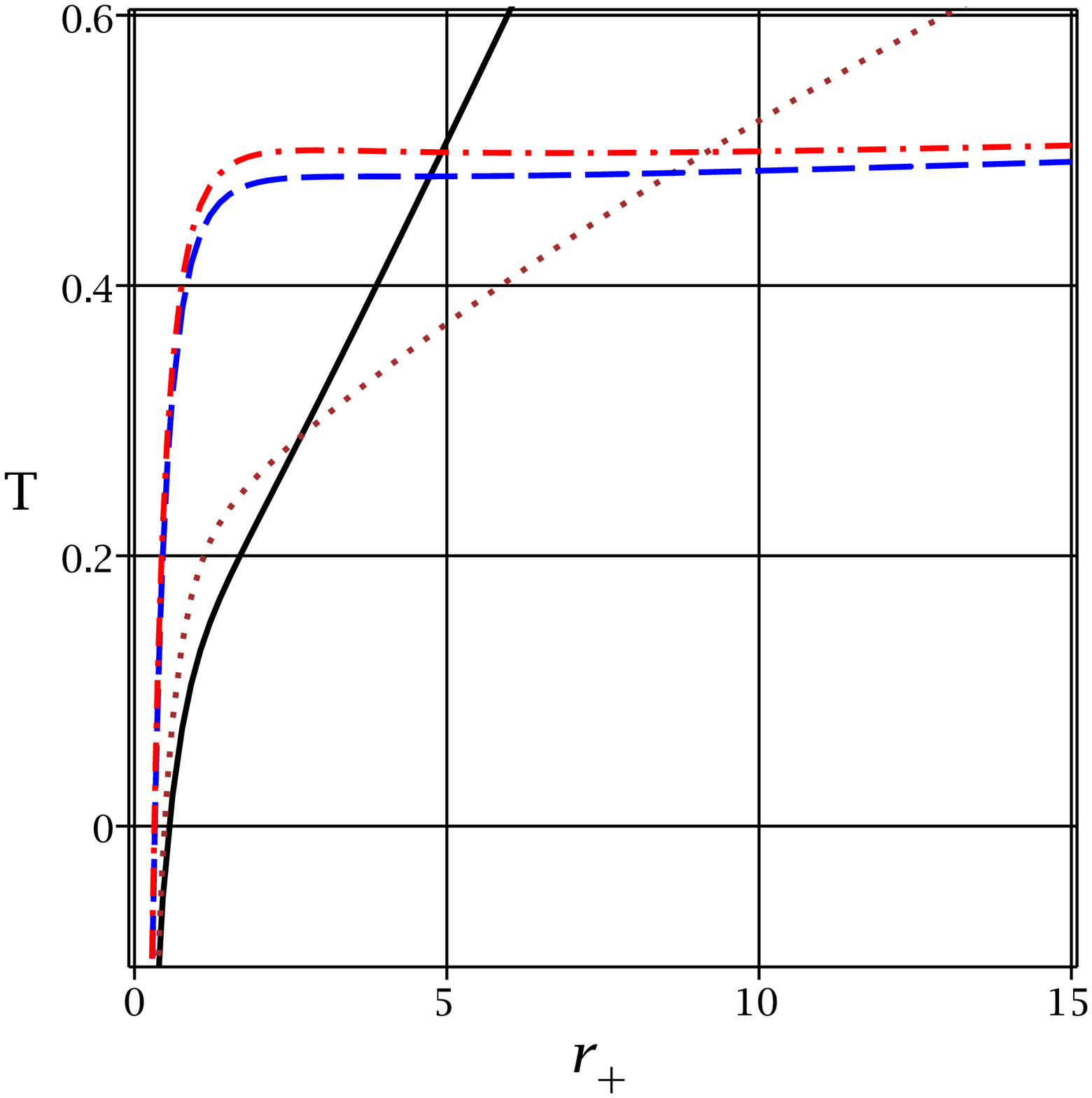} & \epsfxsize=5.5cm \epsffile{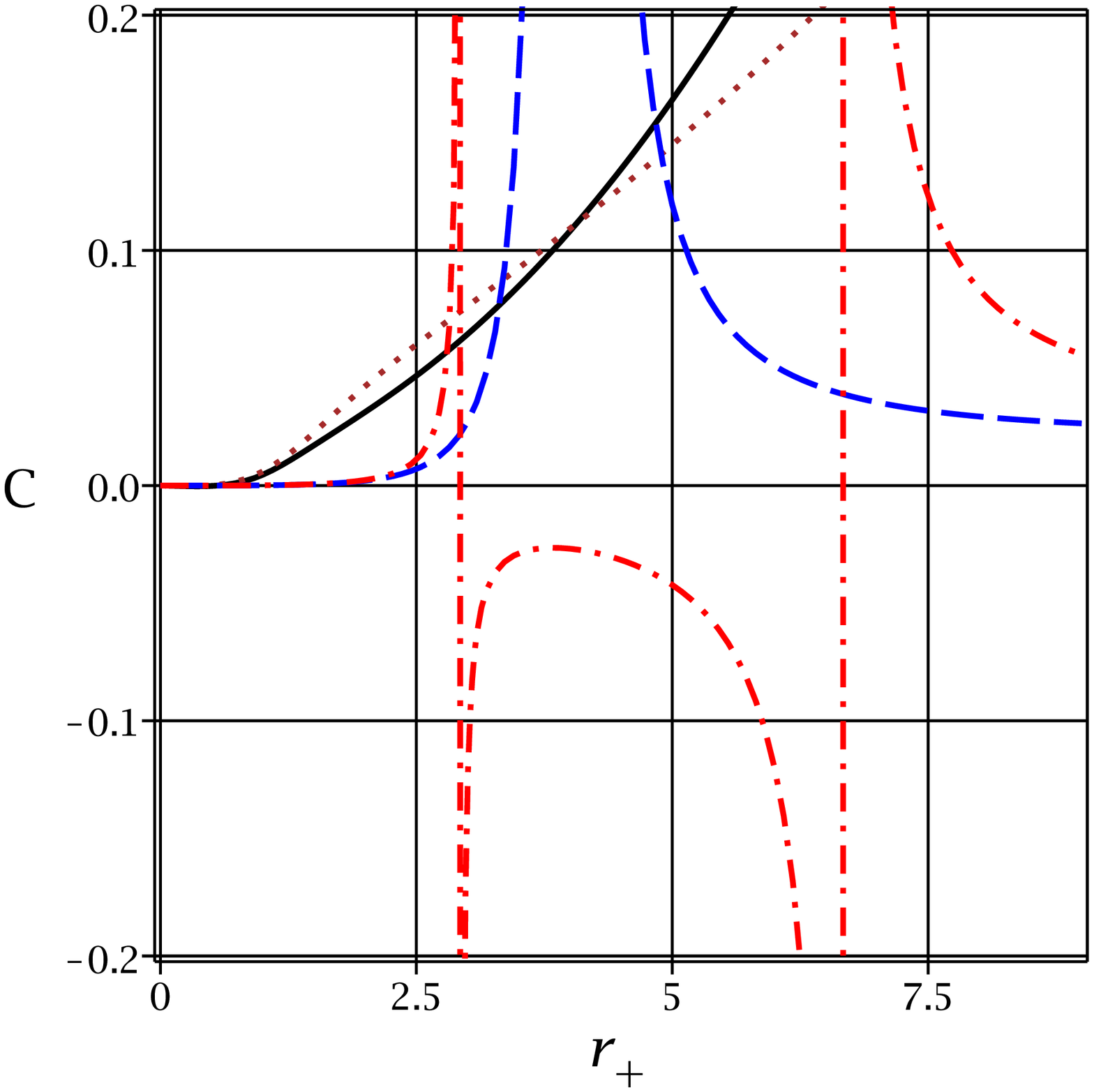} & %
\epsfxsize=5.5cm \epsffile{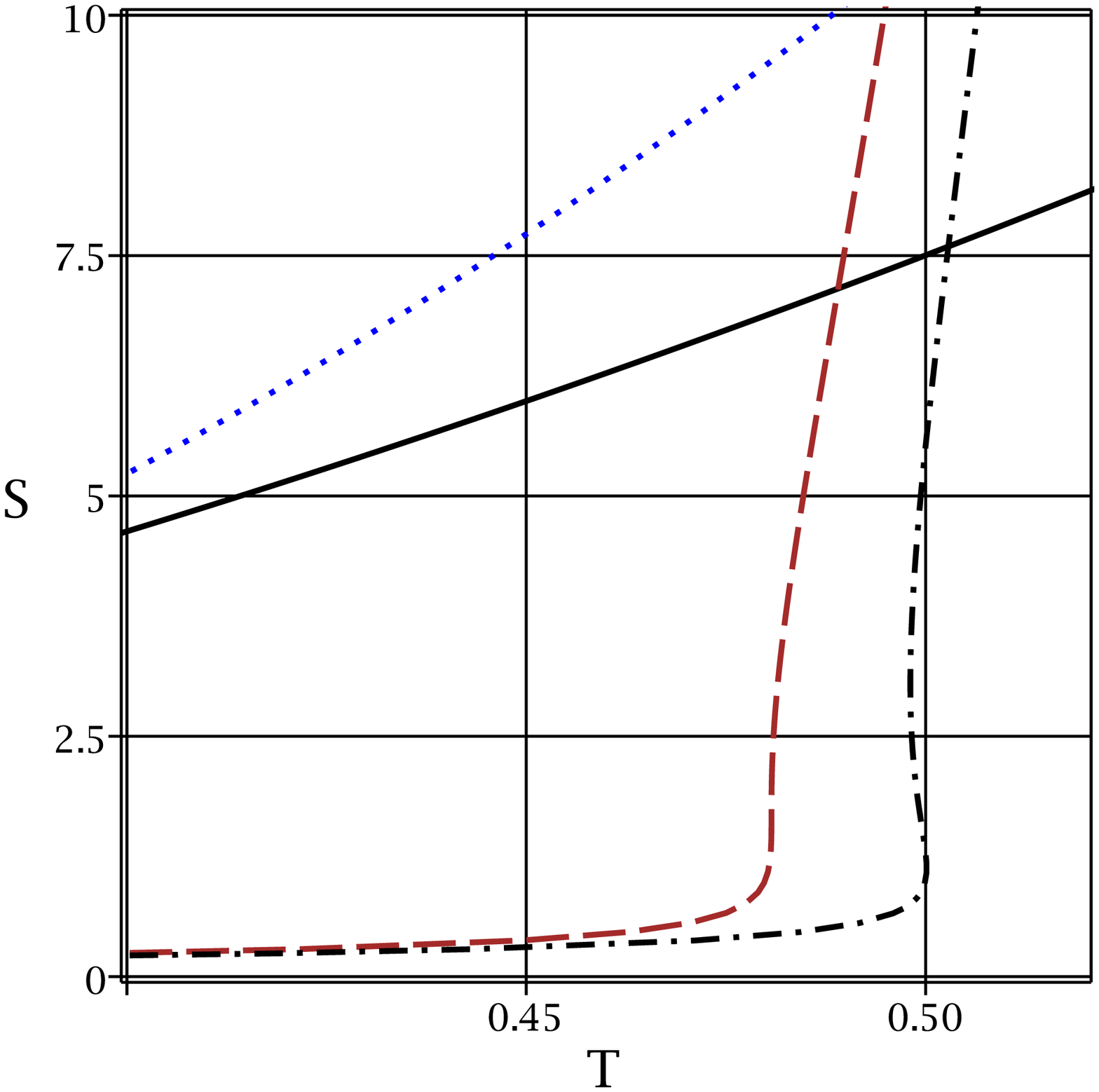}%
\end{array}
$%
\caption{For $k=1$, $m=5$, $\Lambda =-1$, $b=1$, $q=1$, $\protect\beta=1$, $%
f(\protect\varepsilon )=g(\protect\varepsilon )=0.9$, $\protect\alpha =0$
(continuous line), $\protect\alpha =0.5$ (dotted line), $\protect\alpha %
=0.79983$ (dashed line) and $\protect\alpha =0.81$ (dashed-dotted line).
\newline
$T$ (left panel) and $C$ (middle panel) versus $r_{+}$; $S$ versus $T$
(right panel).}
\label{Fig2}
\end{figure*}

\begin{figure*}[tbp]
$%
\begin{array}{ccc}
\epsfxsize=5.5cm \epsffile{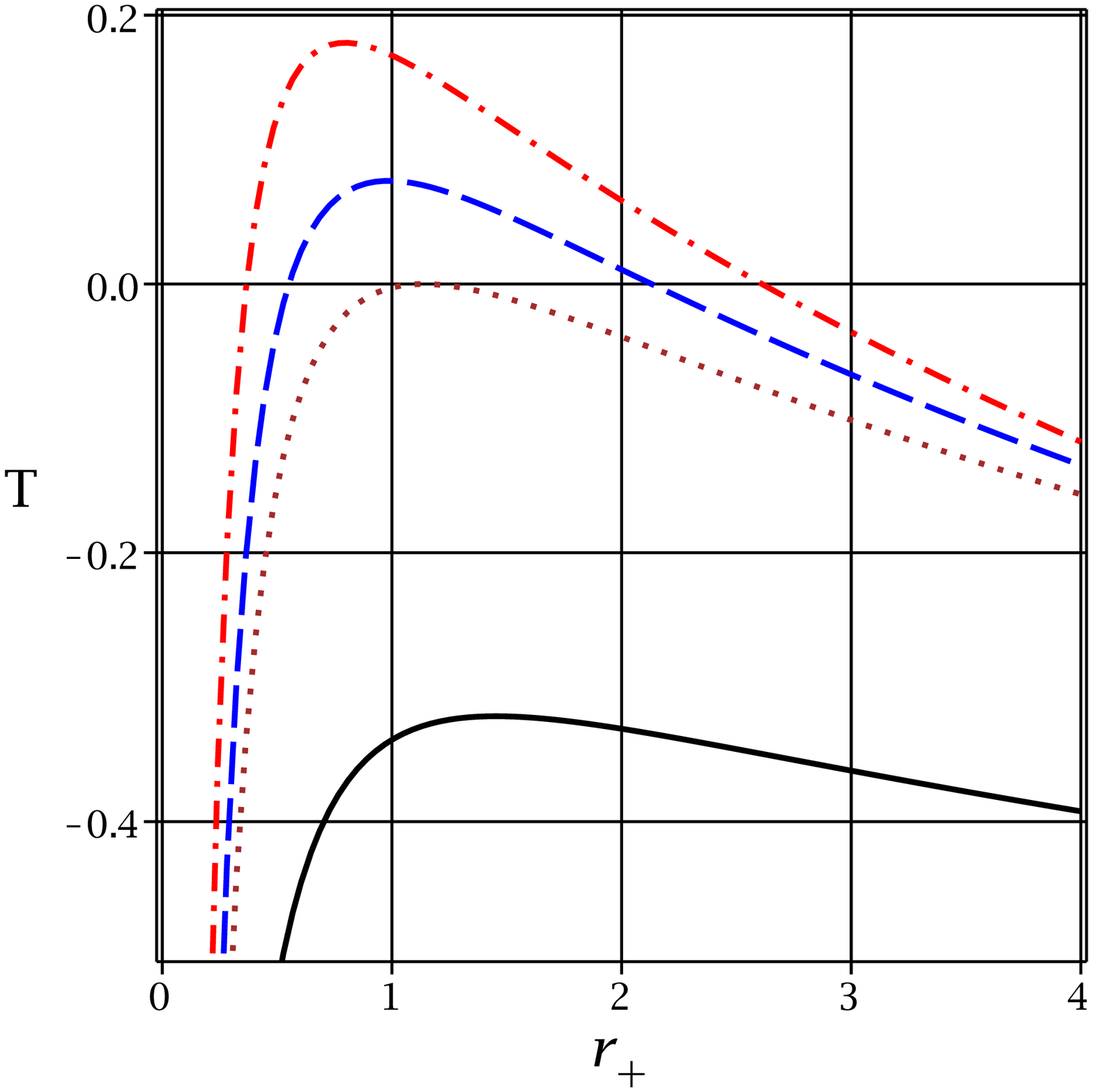} & \epsfxsize=5.5cm \epsffile{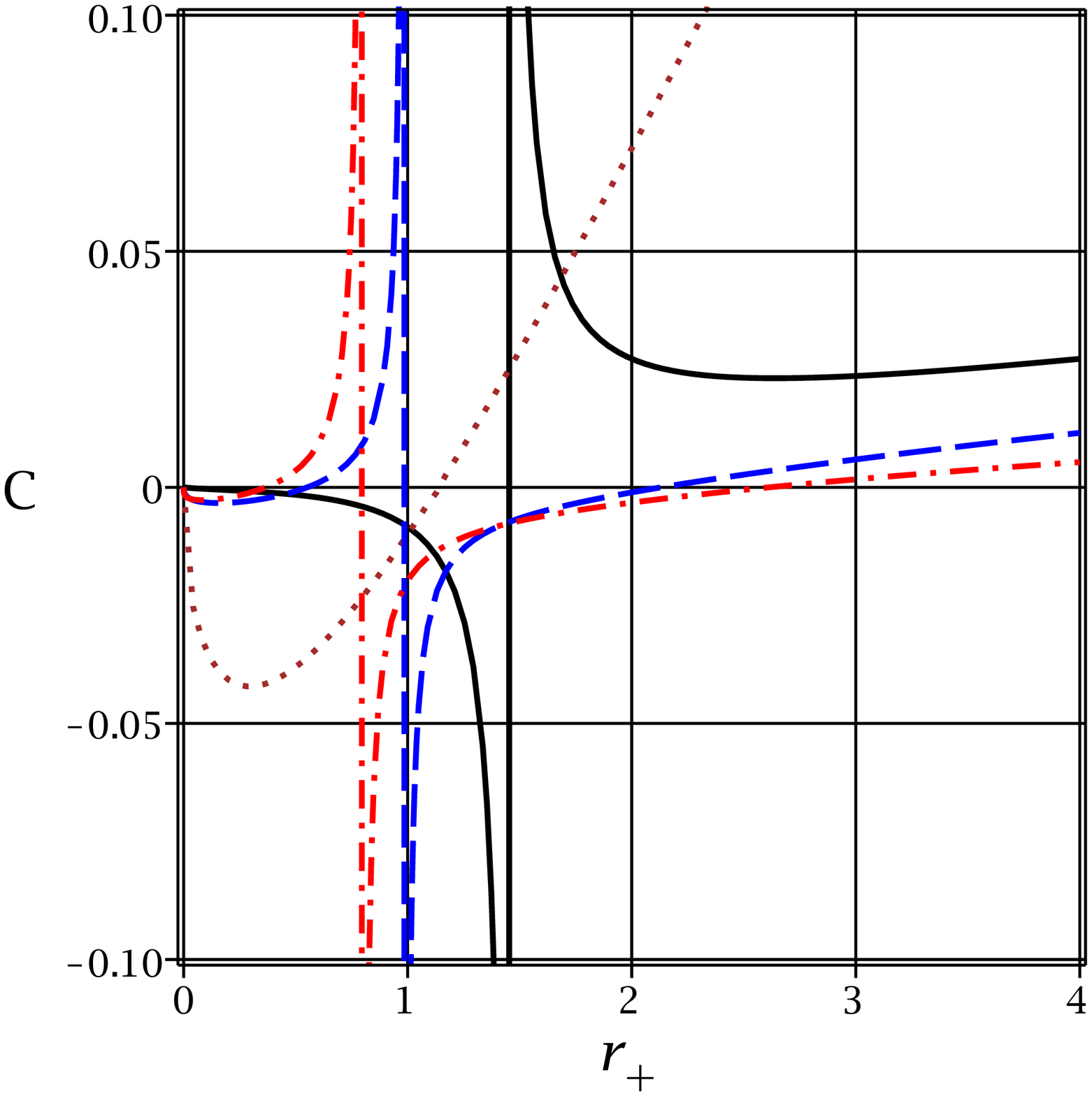} & %
\epsfxsize=5.5cm \epsffile{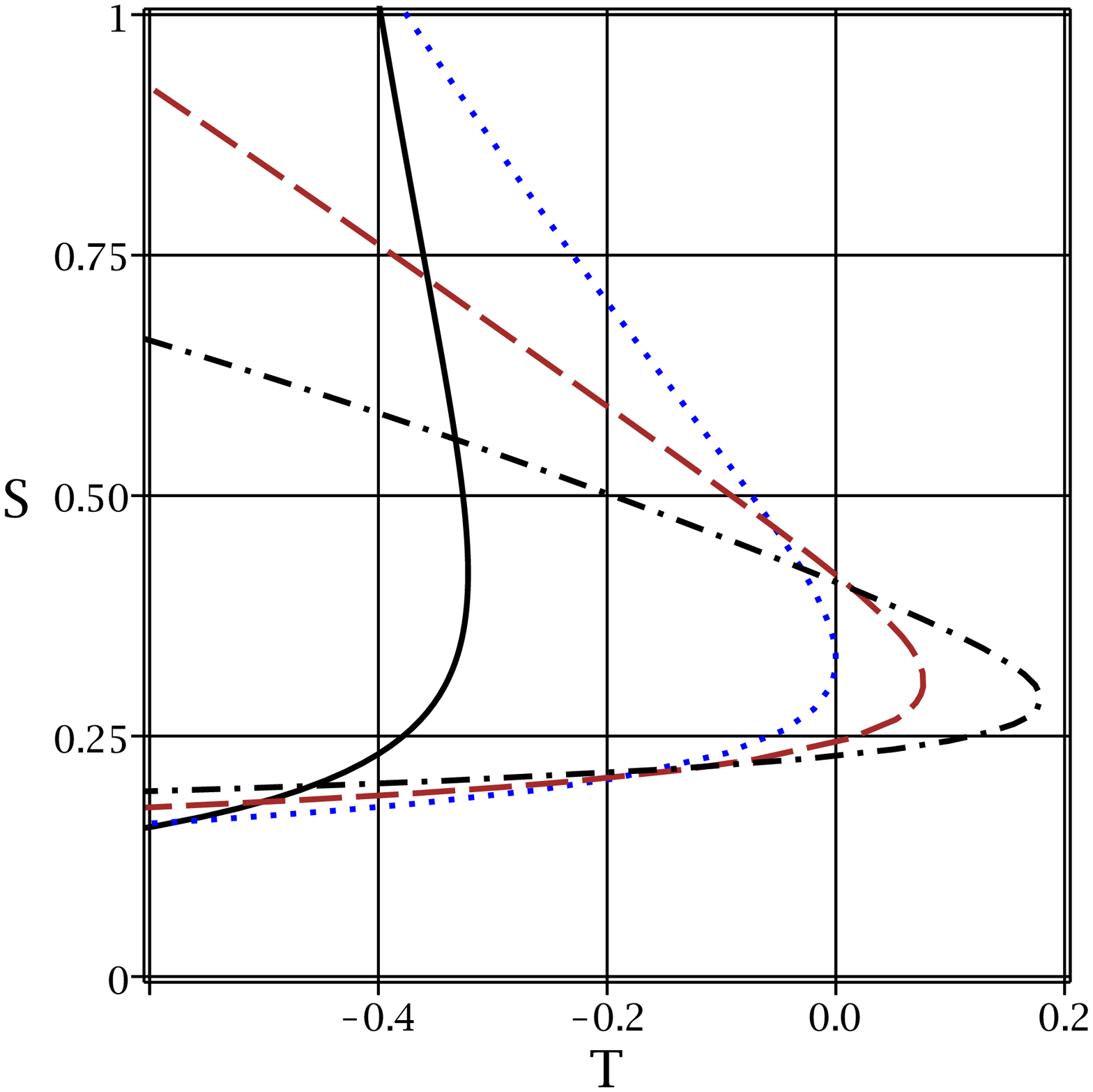}%
\end{array}
$%
\caption{For $k=1$, $m=5$, $\Lambda=-1$, $b=1$, $q=1$, $\protect\beta=1$, $f(%
\protect\varepsilon)=g(\protect\varepsilon)=0.9$, $\protect\alpha=1.2$
(continuous line), $\protect\alpha=1.6941$ (dotted line), $\protect\alpha=2$
(dashed line) and $\protect\alpha=2.4$ (dashed-dotted line). \newline
$T$ (left panel) and $C$ (middle panel) versus $r_{+}$; $S$ versus $T$
(right panel).}
\label{Fig3}
\end{figure*}

Now, in order to investigate the behavior of temperature, we have plotted
various diagrams for variation of dilatonic parameter, $\alpha$. As it was
pointed out, we have divided the effects of dilatonic parameter into two
branches of $\alpha <1$ (left panel of Fig. \ref{Fig2}) and $\alpha >1$
(left panel of Fig. \ref{Fig3}).

First, we investigate $\alpha <1$ case. Evidently, depending on the choices
of dilatonic parameter, temperature could have different properties such as:
I) existence of one root and being increasing function of the horizon radius
(continuous and dotted lines in left panel of Fig. \ref{Fig2}). II)
existence of one root and one extremum (dashed line in left panel of Fig. %
\ref{Fig2}). III) existence of one root and two extrema (dashed-dotted line
in left panel of Fig. \ref{Fig2}). Evidently, for small values of the
dilatonic parameter, the contribution of scalar field is not significant and
general behavior of temperature is similar to the absence of this parameter.
Increasing the dilatonic parameter leads to formation of an extremum.
Further increment leads to existence of two extrema: a maximum and then a
minimum in which maximum is formed in smaller horizon radius comparing to
minimum. The extrema in temperature are matched with divergencies in the
heat capacity. Therefore, one can conclude that in the case of $\alpha <1$,
increasing the dilatonic parameter leads to introduction of critical
behavior in thermodynamical structure of the black holes. Later in studying
the heat capacity, we will give more details regarding the types of critical
behavior that could exist for these black holes.

Let us now turn our attention to the case of $\alpha >1$. Interestingly, for
small values of the dilatonic parameter, the temperature is negative valued
everywhere (continuous line in left panel of Fig. \ref{Fig3}). This
indicates that our solutions are thermodynamically non-physical. It is
worthwhile to mention that for this case, temperature enjoys a maximum in
its structure. The maximum value of temperature is an increasing function of
dilatonic parameter. For specific value of dilatonic parameter, one can find
a root for temperature (dotted line in left panel of Fig. \ref{Fig3}). The
root is an extreme point, but then again, we should point it out that except
for the root, temperature is negative everywhere. Increasing dilatonic
parameter more than this specific value leads to formation of two roots for
the temperature (dashed and dashed-dotted lines in left panel of Fig. \ref%
{Fig3}). The maximum is located between these roots, which shows that the
physical black holes could only be observed for medium black holes. Whereas,
for small and large black holes, temperature is negative valued and
solutions are thermally non-physical.

The behavior that we observed in plotted diagrams for $\alpha <1$ case
actually shows the existence of subcritical isobars. Presence of subcritical
isobar, so far has been reported only for AdS black holes. On the other
hand, the behavior that we observed for temperature in $\alpha >1$ case was
similar to the one that previously was reported for dS black holes. These
two specific properties confirms a very important result regarding the
dilatonic parameter: for $\alpha <1$ case, thermodynamical behavior or
temperature of black holes is AdS like while for $\alpha>1$ case, this
behavior is dS like. This indicates that in general, for $\alpha <1$, the
black holes have AdS like behavior while for the $\alpha >1$ the behavior is
dS like. If one is interested to study the AdS/CFT duality in the context of
these black holes, the valid branch is where $\alpha <1$.

Now, we turn our attention to the heat capacity which contains information
regarding thermal stability and possible phase transition points. Unlike
temperature, the heat capacity could be correction dependent. In other
words, considering the first order correction to the entropy, the heat
capacity would be modified. Interestingly, the effects of correction could
be observed in the numerator of heat capacity while its denominator is
independent of it. This confirms two important things regarding the effects
of first order correction: I) the phase transition points that are realized
through divergencies in the heat capacity are not affected by the first
order correction. In other words, the first order correction has no effect
on phase transition points, II) in usual black holes without the first order
correction, the heat capacity and temperature share same roots. But since
the numerator of heat capacity in the presence of first order correction is
modified, temperature and heat capacity no longer share the same roots. This
modifies the conditions regarding thermal stability of the solutions.

\begin{figure*}[tbp]
$%
\begin{array}{ccc}
\epsfxsize=5.5cm \epsffile{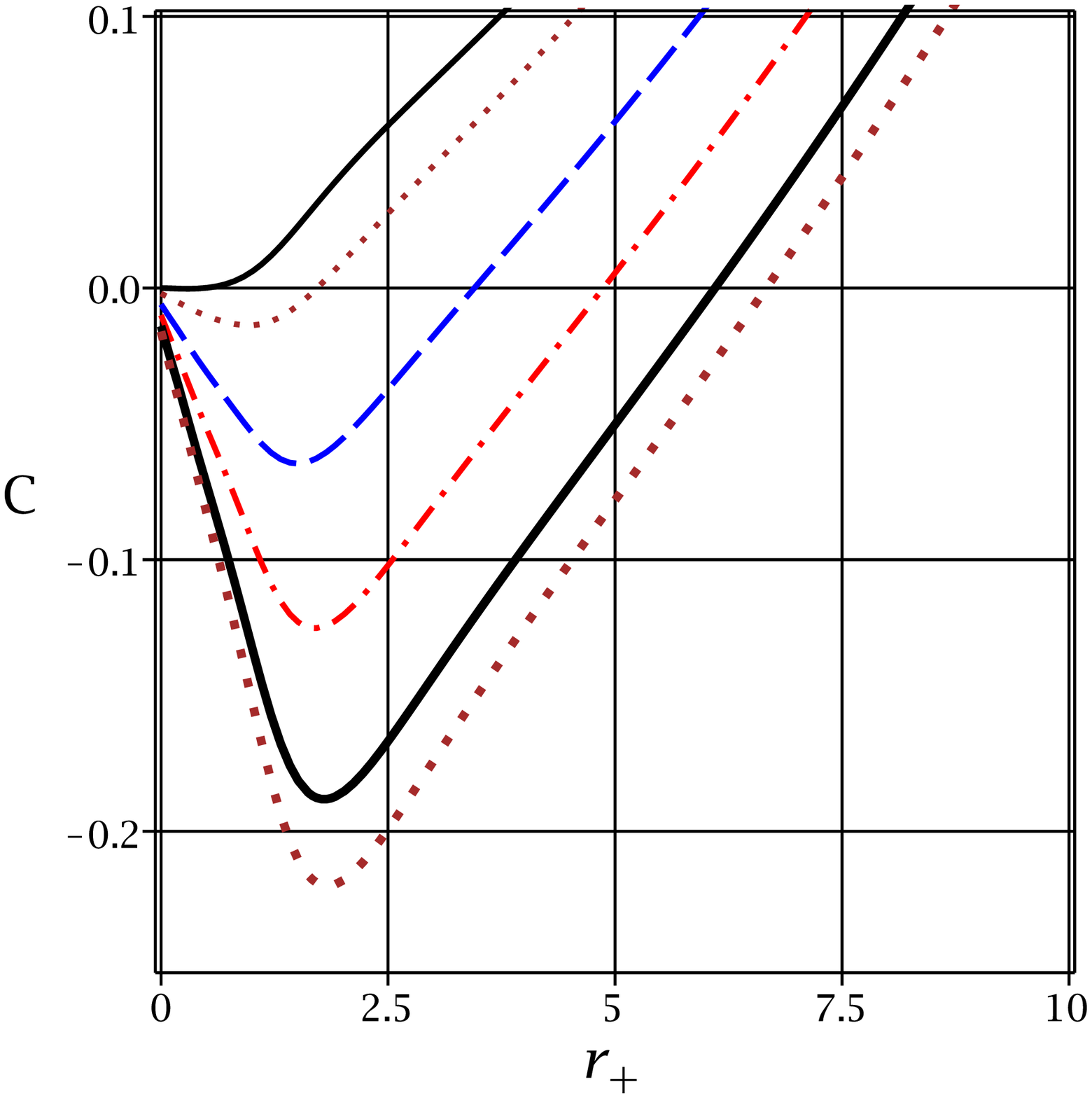} & \epsfxsize=5.5cm \epsffile{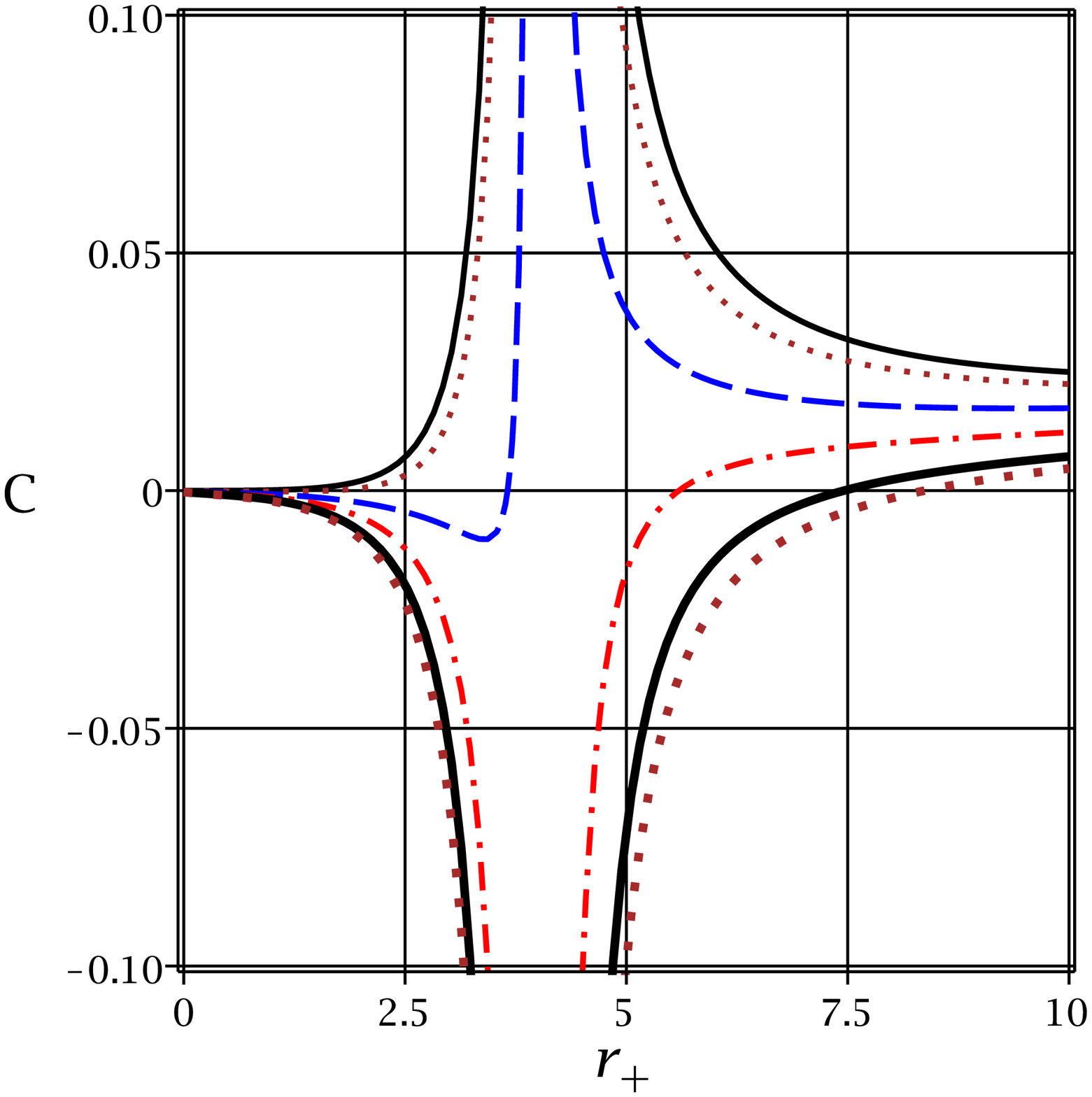}
& \epsfxsize=5.5cm \epsffile{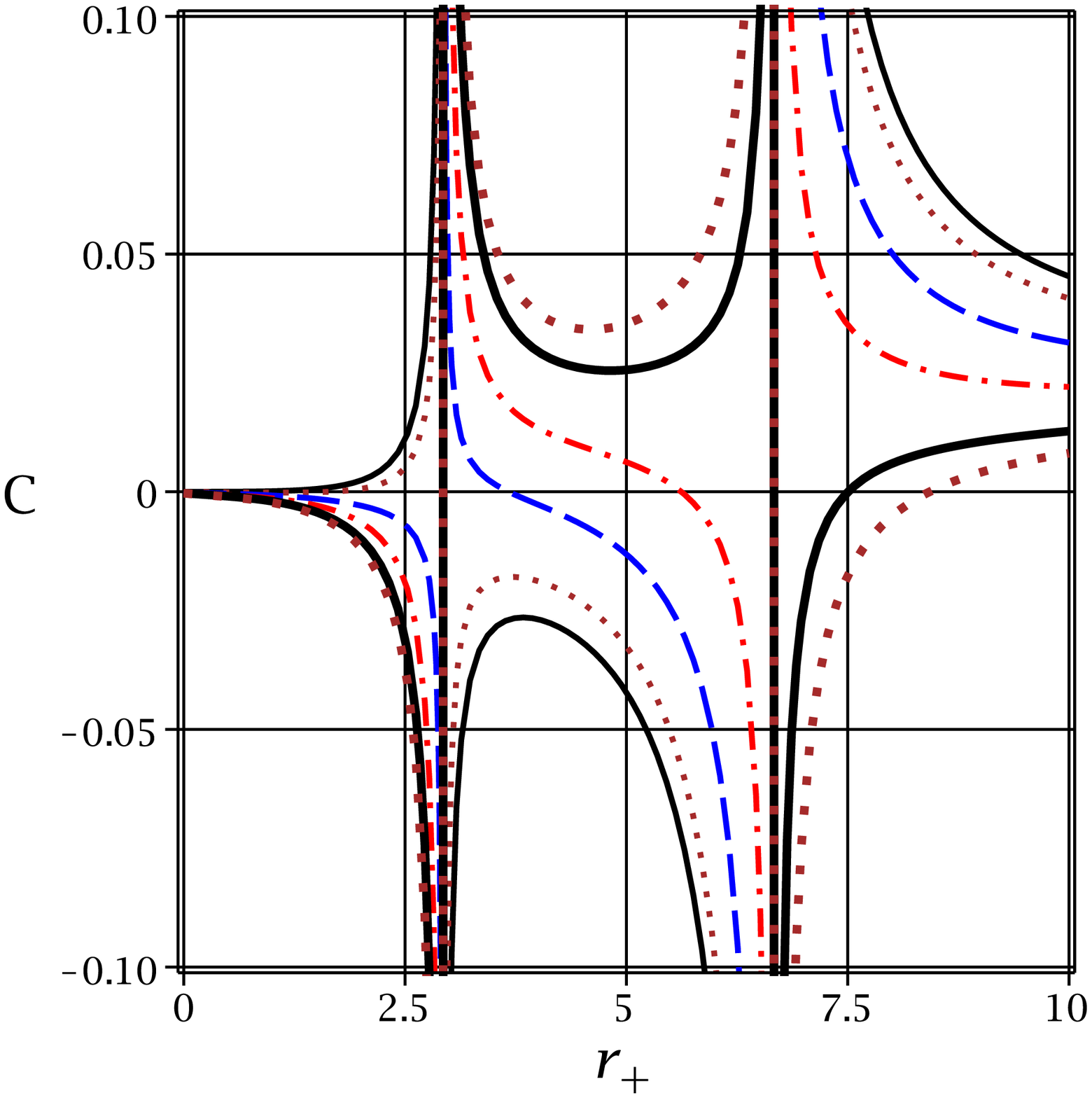} \\
\epsfxsize=5.5cm \epsffile{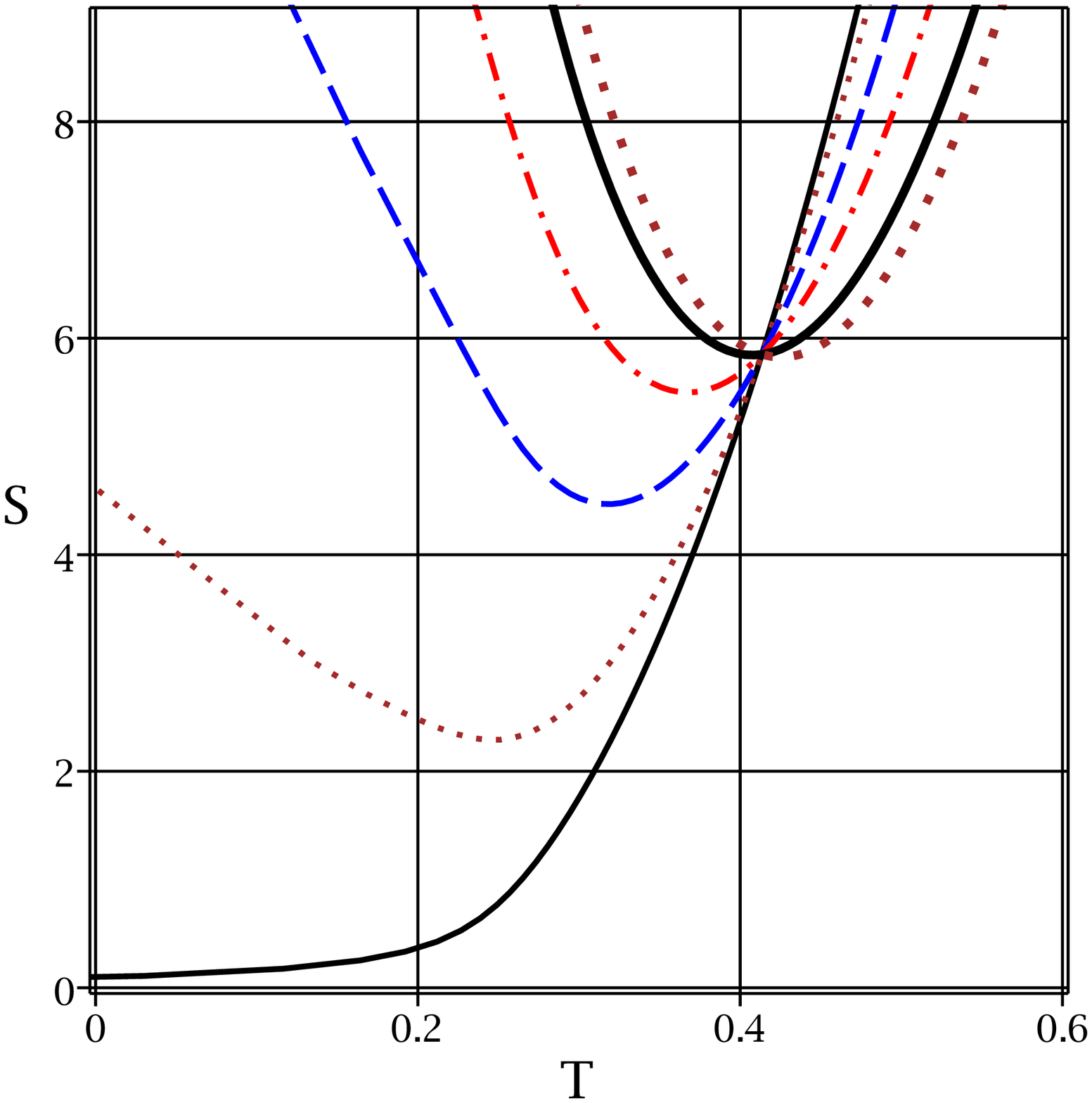} & \epsfxsize=5.5cm \epsffile{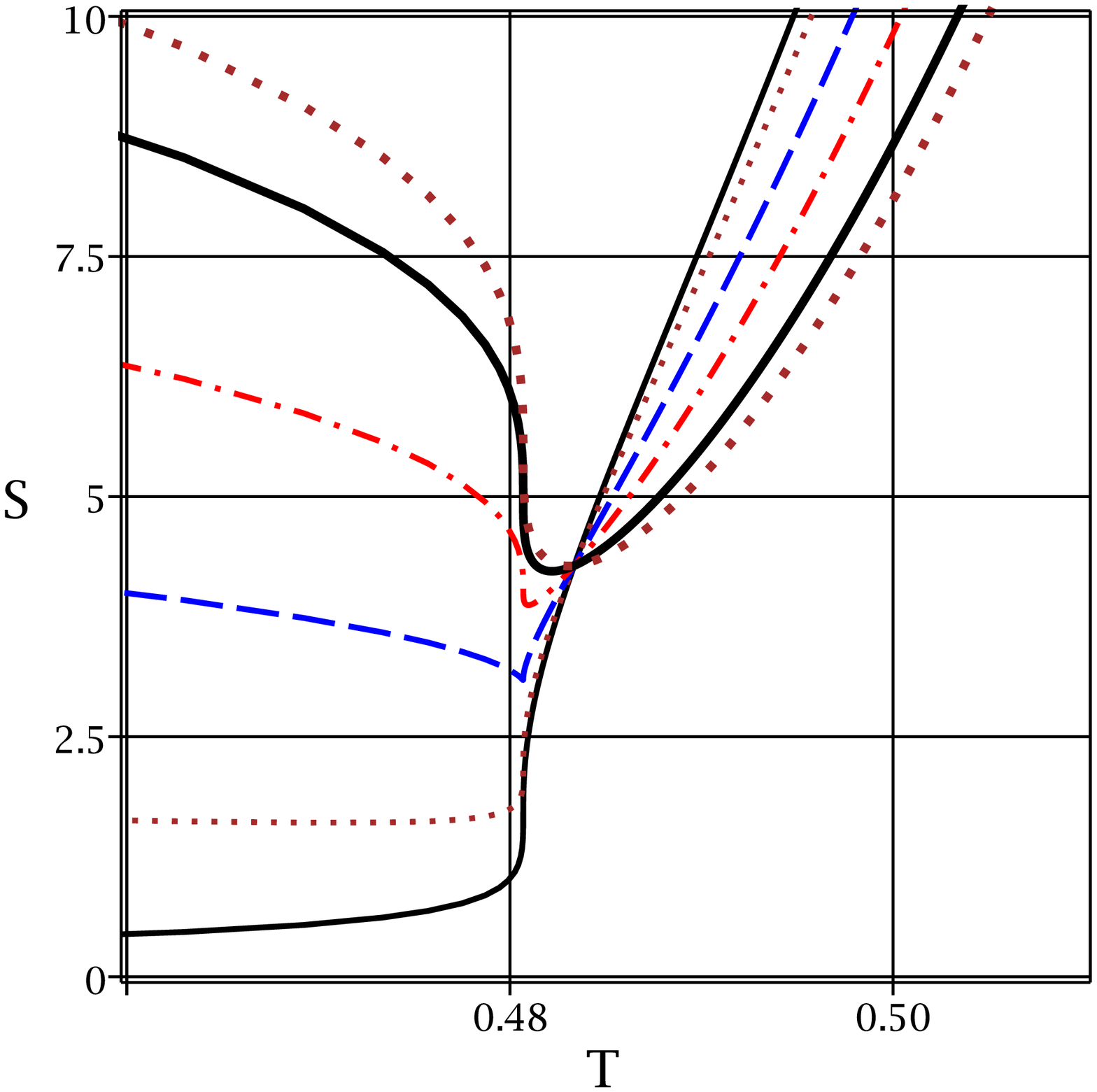}
& \epsfxsize=5.5cm \epsffile{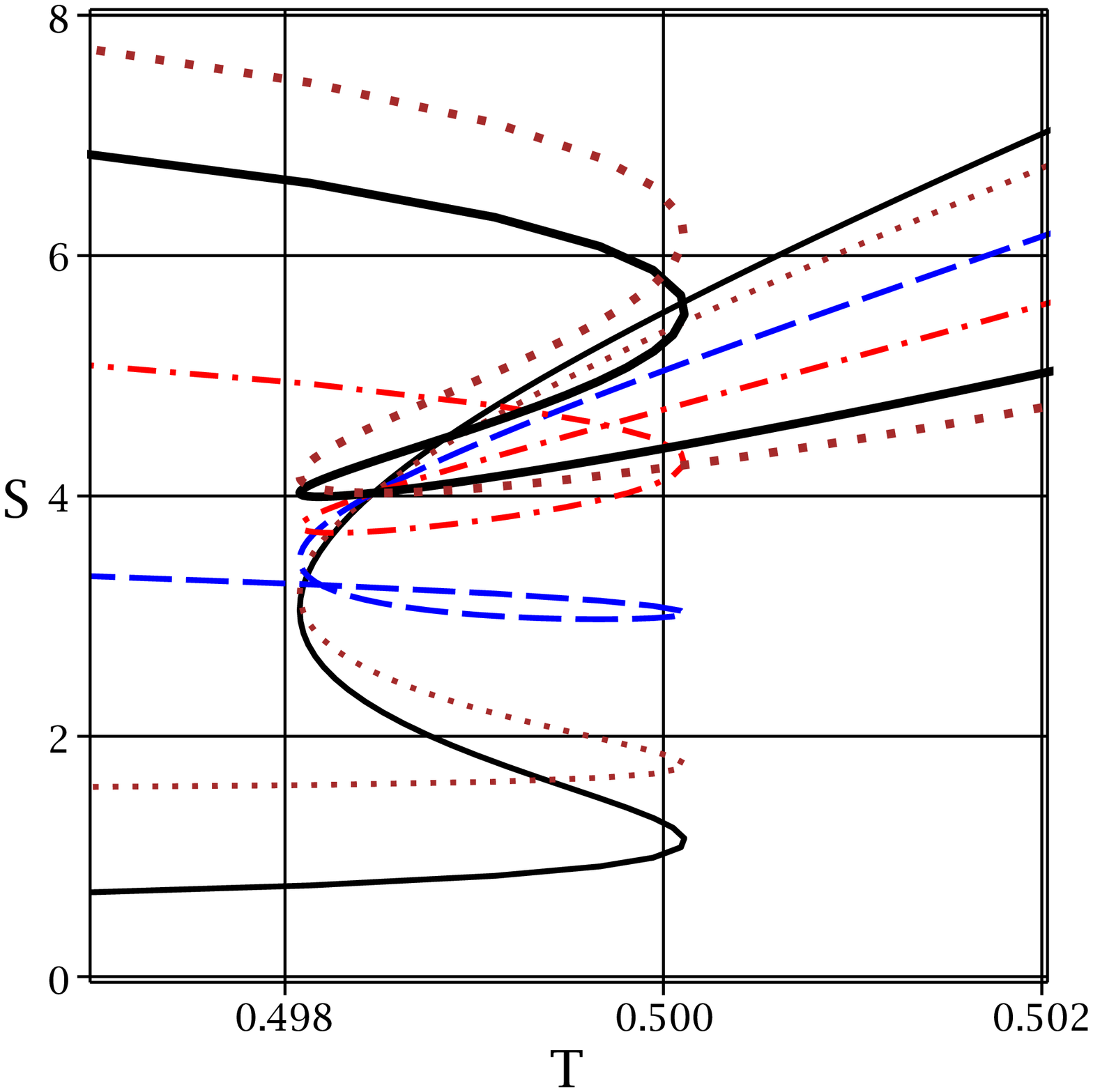}%
\end{array}
$%
\caption{$C$ versus $r_{+}$ (up panels) and $S$ versus $T$ (down panels) for
$k=1$, $m=5$, $\Lambda=-1$, $b=1$, $q=1$, $\protect\beta=1$, $f(\protect%
\varepsilon)=g(\protect\varepsilon)=0.9$, $\protect\zeta=0$ (continuous
line), $\protect\zeta=1$ (dotted line), $\protect\zeta=3$ (dashed line), $%
\protect\zeta=5$ (dashed-dotted line), $\protect\zeta=7$ (bold continuous
line) and $\protect\zeta=8$ (bold dotted line). \newline
Left panel: $\protect\alpha=0.5$; middle panel: $\protect\alpha=0.79883$;
right panel: $\protect\alpha=0.81$.}
\label{Fig4}
\end{figure*}

We recall that in the case of $\alpha<1$, the behavior is AdS like. For this
case, there exists a critical value for dilatonic parameter, say $\alpha_{c}$
($c$ stands for "critical") which could be used to divide possible scenarios
available for the heat capacity. For $0<\alpha <\alpha _{c}$, the effects of
dilatonic gravity is significant on the place of root of heat capacity.
Before the root, the heat capacity is negative and solutions are thermally
unstable while after it, the opposite is seen and solutions are thermally
stable (see continuous and dotted lines in middle panel of Fig. \ref{Fig2}).
This shows that for small values of the dilatonic parameter, region of
stability is modified. At the critical value ($\alpha =\alpha _{c}$), the
heat capacity acquires a divergency which is interpreted as critical
behavior (dashed line in middle panel of Fig. \ref{Fig2}). The sign of heat
capacity around this divergence point is positive which indicates that the
critical characteristic takes place between two stable phases, as expected.
In this case, the heat capacity enjoys a root as well, and it is located
before the divergency. Finally, for $\alpha _{c}<\alpha <1$, the heat
capacity has two divergencies in its structure (dashed-dotted line in middle
panel of Fig. \ref{Fig2}). Between the divergencies, the sign of heat
capacity is negative which shows that solutions are thermally unstable. This
indicates that there is a phase transition taking place between two
divergencies. Therefore, the only stable phases provided for the black holes
in this case are small and large black holes while the medium black holes
suffers thermal instability. It is worthwhile to mention that before the
smaller divergency, there also exists a root for the heat capacity. Before
this root, the temperature is negative, therefore solutions are
non-physical. The behavior that we observed for the heat capacity here
completely matches to the one that was observed for the heat capacity of AdS
black holes (see appendix for more details). This confirms that thermal
stability structure of these black holes in case of $\alpha <1$ is same as
AdS black holes. This provides us with further proof to recognize the branch
$\alpha <1$ as AdS spacetime.

\begin{figure*}[tbp]
$%
\begin{array}{ccc}
\epsfxsize=5.5cm \epsffile{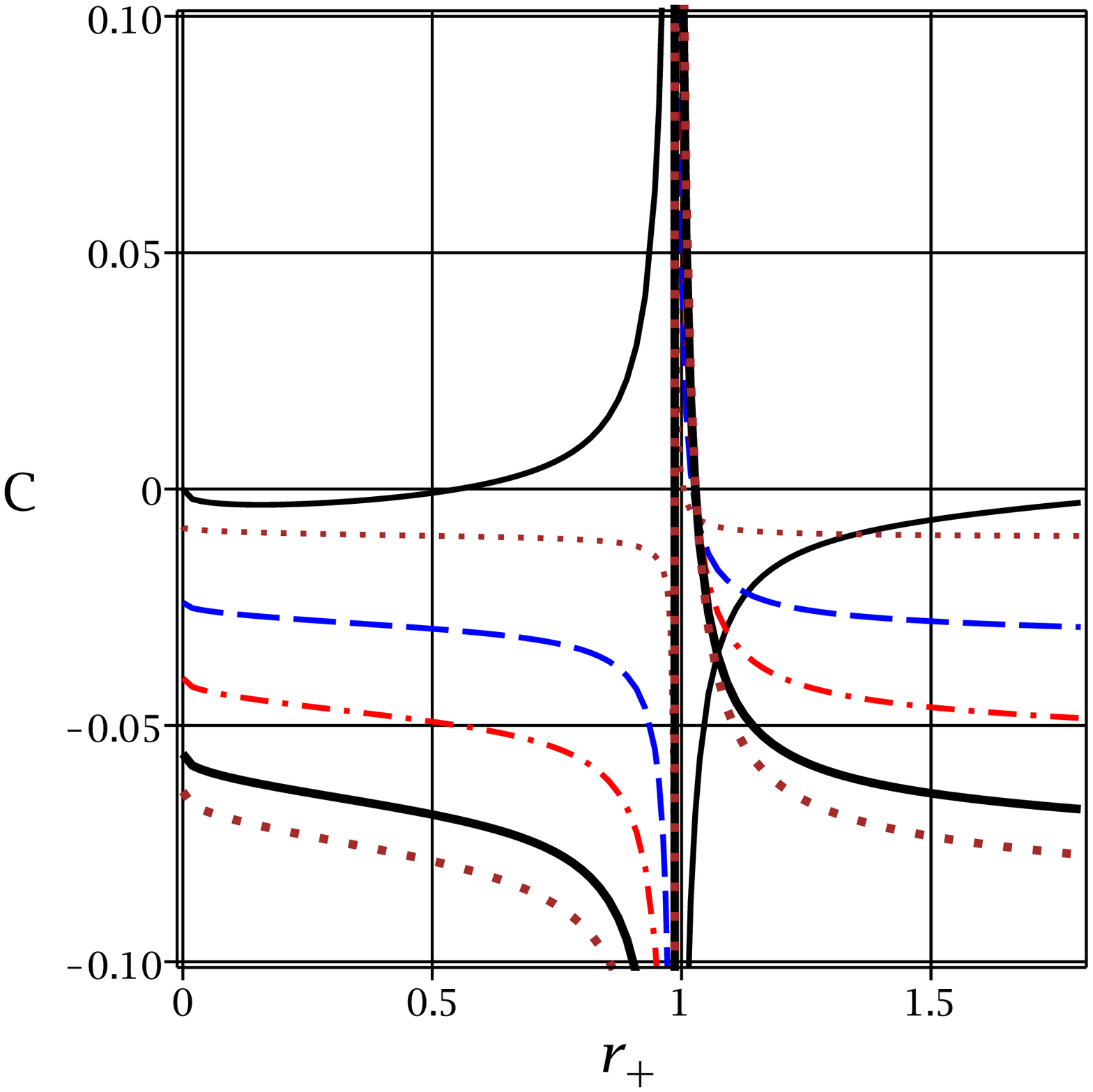} & \epsfxsize=5.5cm \epsffile{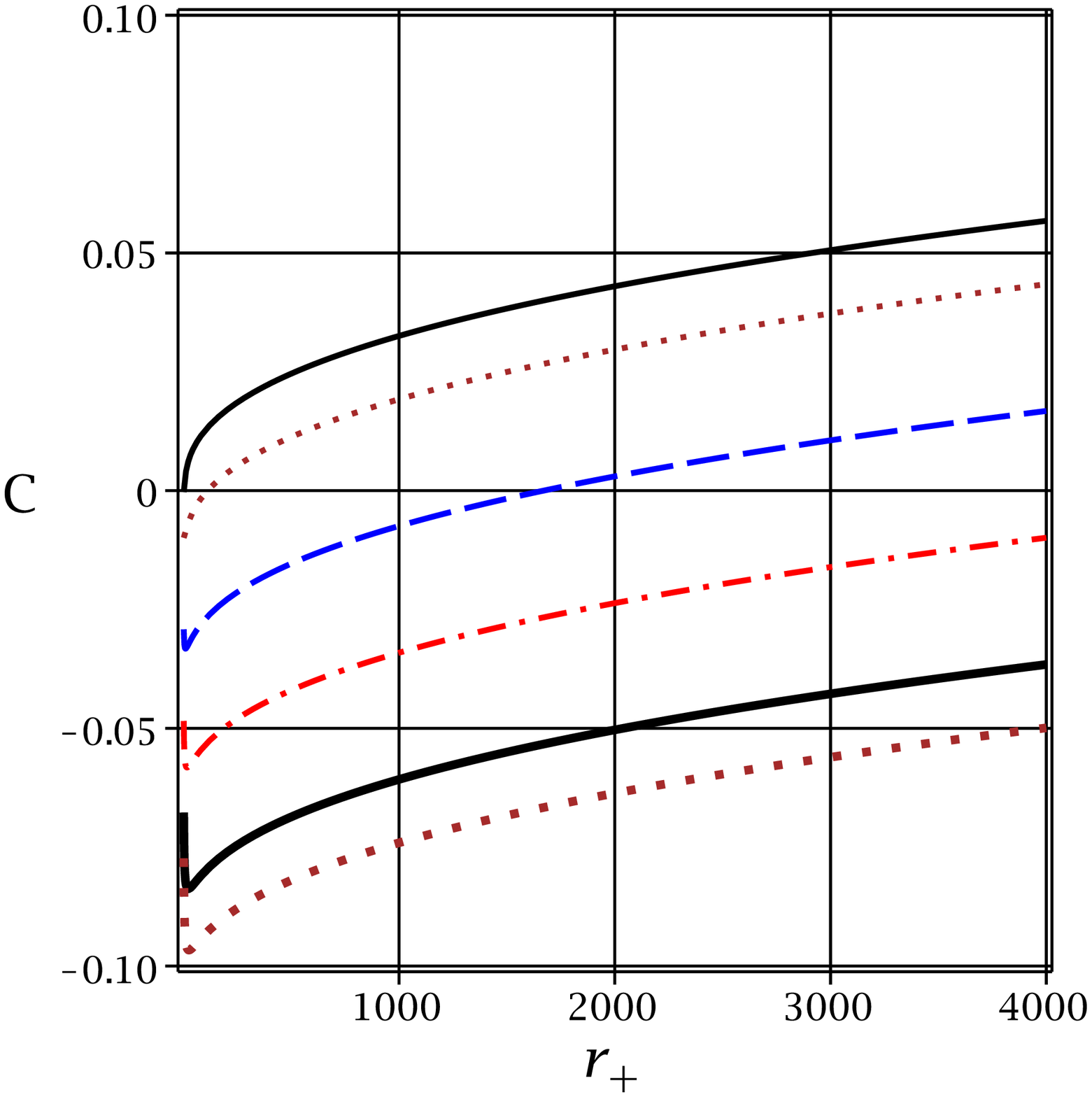}
&  \\
\epsfxsize=5.5cm \epsffile{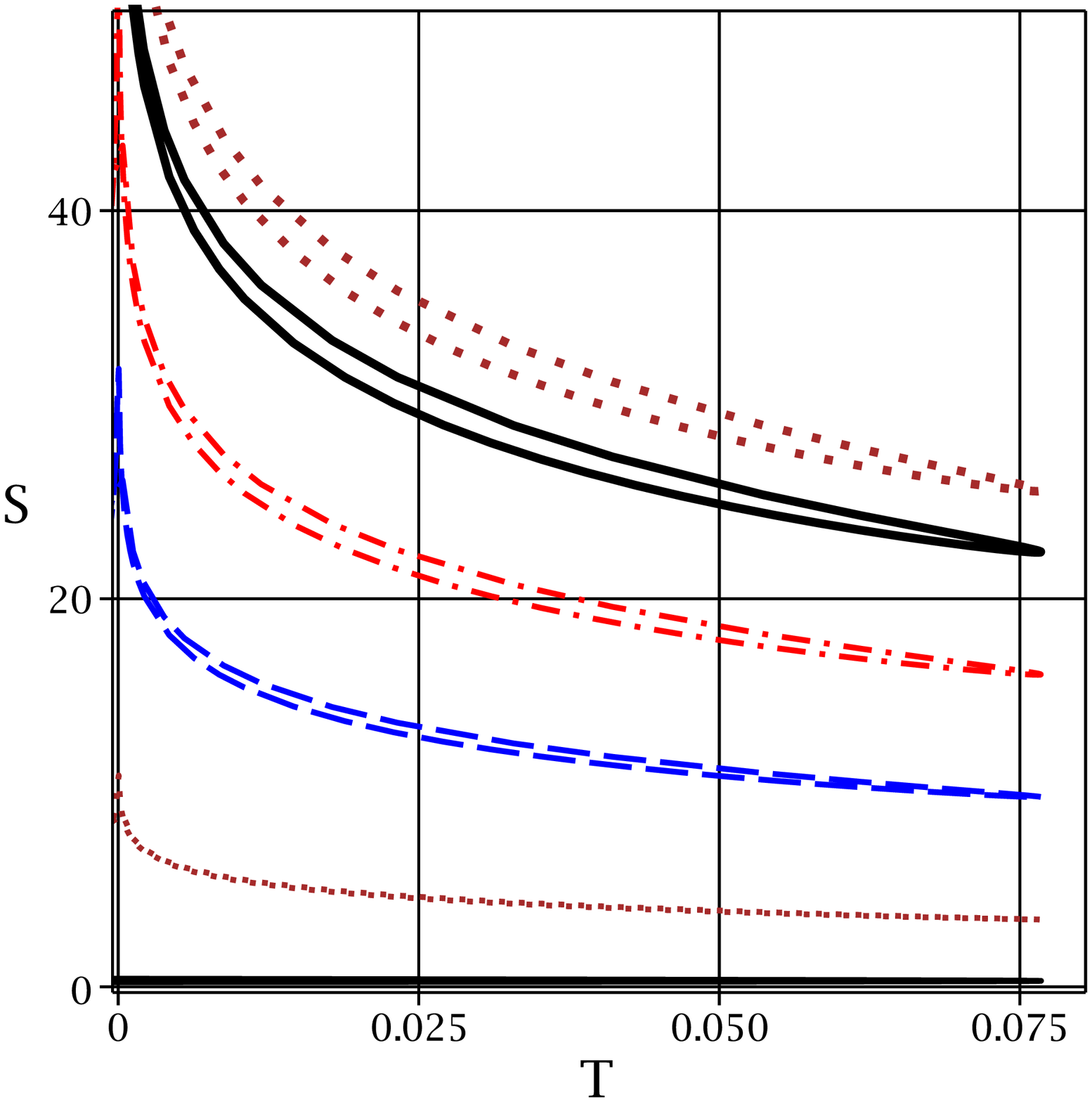} & \epsfxsize=5.5cm \epsffile{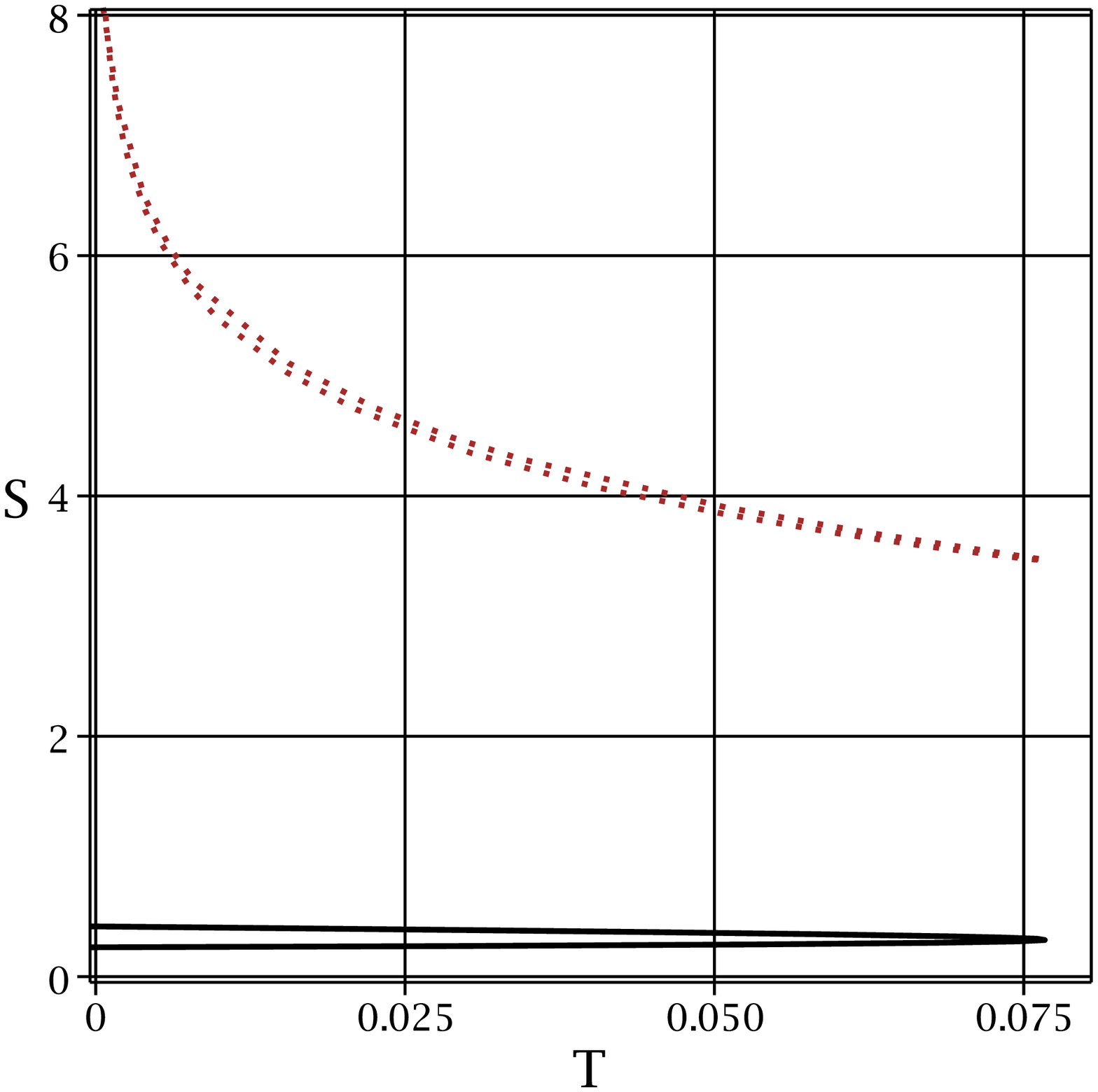}
&
\end{array}
$%
\caption{$C$ versus $r_{+}$ (up panels) and $S$ versus $T$ (down panels) for
$k=1$, $m=5$, $\Lambda=-1$, $b=1$, $q=1$, $\protect\beta=1$, $f(\protect%
\varepsilon)=g(\protect\varepsilon)=0.9$, $\protect\alpha=2$, $\protect\zeta%
=0$ (continuous line), $\protect\zeta=1$ (dotted line), $\protect\zeta=3$
(dashed line), $\protect\zeta=5$ (dashed-dotted line), $\protect\zeta=7$
(bold continuous line) and $\protect\zeta=8$ (bold dotted line).}
\label{Fig5}
\end{figure*}


In case of $\alpha >1$, the general behavior in temperature was dS like.
Here, it is possible to divide the general behavior of the heat capacity
into three groups by a specific value of $\alpha$, say $\alpha_{er}$ ($er$
stands for extreme root). For $1<\alpha <\alpha _{er}$, the heat capacity
has only one divergency in which the heat capacity switches from negative to
positive (continuous line in middle panel of Fig. \ref{Fig3}). This case has
negative temperature everywhere. Therefore, although the heat capacity
signals the existence of stable state, the negative temperature shows that
no physical solution exists in this case. This highlights the importance of
studying the temperature alongside of the heat capacity to separate physical
solutions from non-physical ones. For $\alpha =\alpha _{er}$, interestingly
no divergency is observed for the heat capacity, although the temperature
gives us the detail of its existence by having a maximum (dotted line in
left and middle panels of Fig. \ref{Fig3}). In this case, the temperature is
negative valued everywhere except at its root which is an extreme one. But
the heat capacity in this case shows the existence of only one root which
after it, the heat capacity is positive valued. The absence of divergency in
heat capacity is due to the fact that extremum and root of the temperature
are identical. Considering the relation for obtaining heat capacity (\ref%
{heat}), it is obvious that no divergency could be observed for the heat
capacity in this case. Finally, for $\alpha _{er}<\alpha $, the heat
capacity enjoys two roots with a divergency located between them (dashed and
dashed-dotted lines in middle panel of Fig. \ref{Fig3}). Between smaller
root and divergency, and after larger root, the heat capacity is positive.
Whereas, before smaller root, and between divergency and larger root, the
heat capacity is negative and solutions are thermally unstable. But we
should remind the reader that the only physical region for this case is
between two roots (positive $T$). Therefore, there is one physical stable
phase and a physical unstable one around divergency and small black holes
are stable. The behaviors that we have observed for the heat capacity in
this case ($\alpha>1$) is exactly the same as the one that was observed for
dS case (see appendix for more details). Therefore, this confirms the
analogy of consideration of $\alpha >1$ as dS case.

Now, we give more details regarding the effects of first order correction on
thermodynamical behavior of the solutions. To do so, we have considered
different behaviors that were reported for the heat capacity before and
study the effects of variation of the first order parameter, $\zeta$.

\begin{figure*}[tbp]
$%
\begin{array}{cc}
\epsfxsize=5.5cm \epsffile{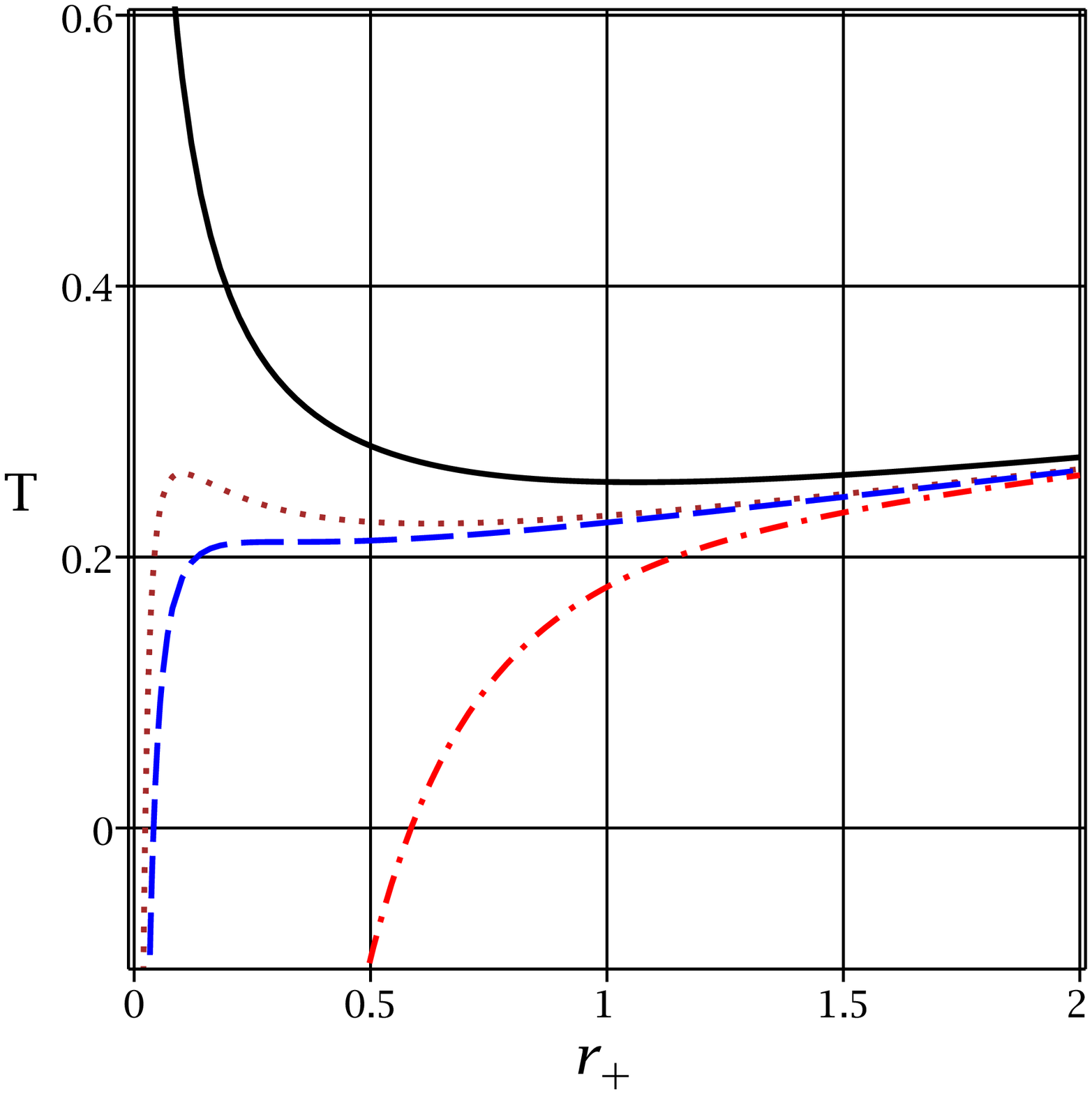} & \epsfxsize=5.5cm \epsffile{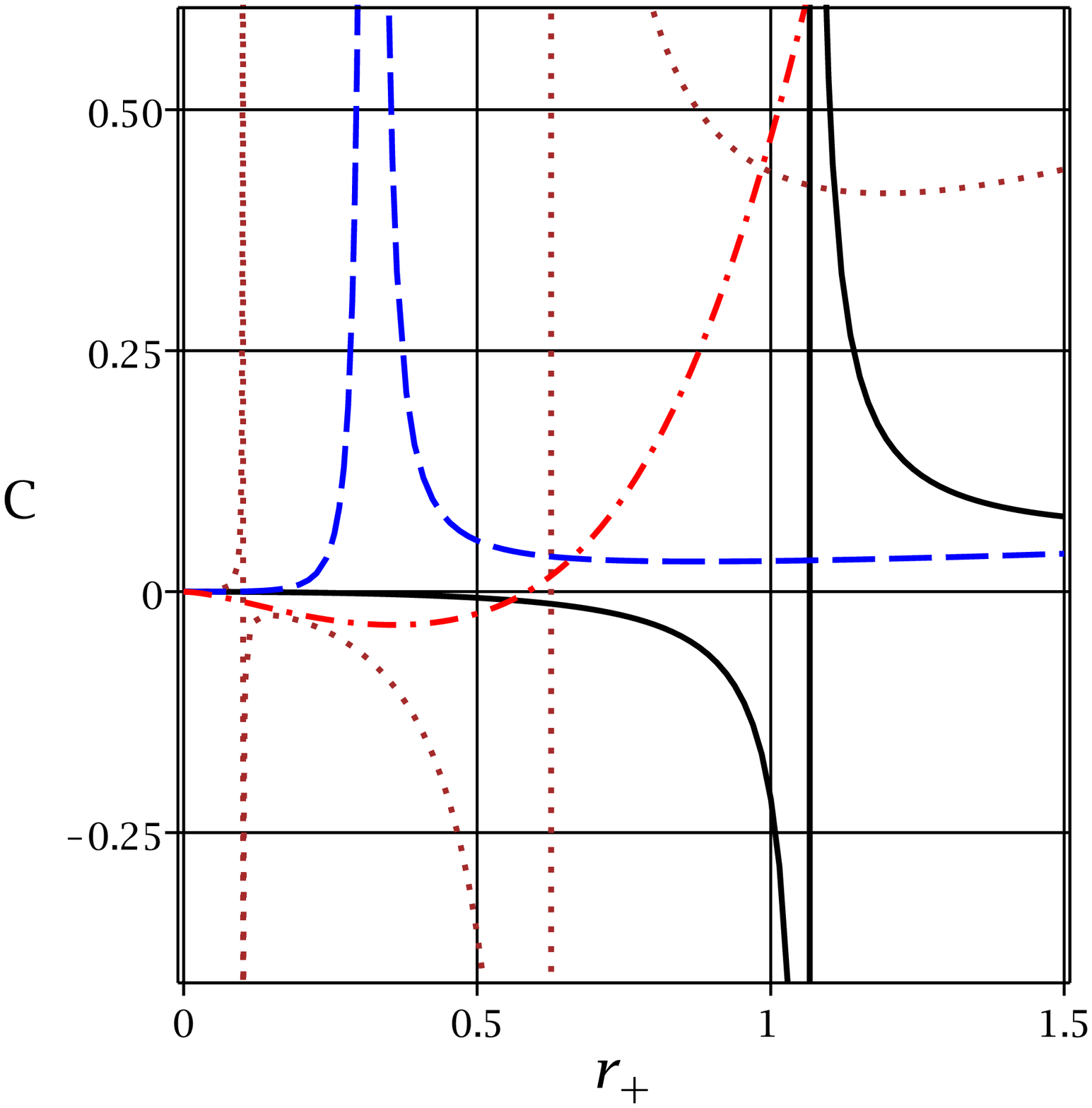}
\\
\epsfxsize=5.5cm \epsffile{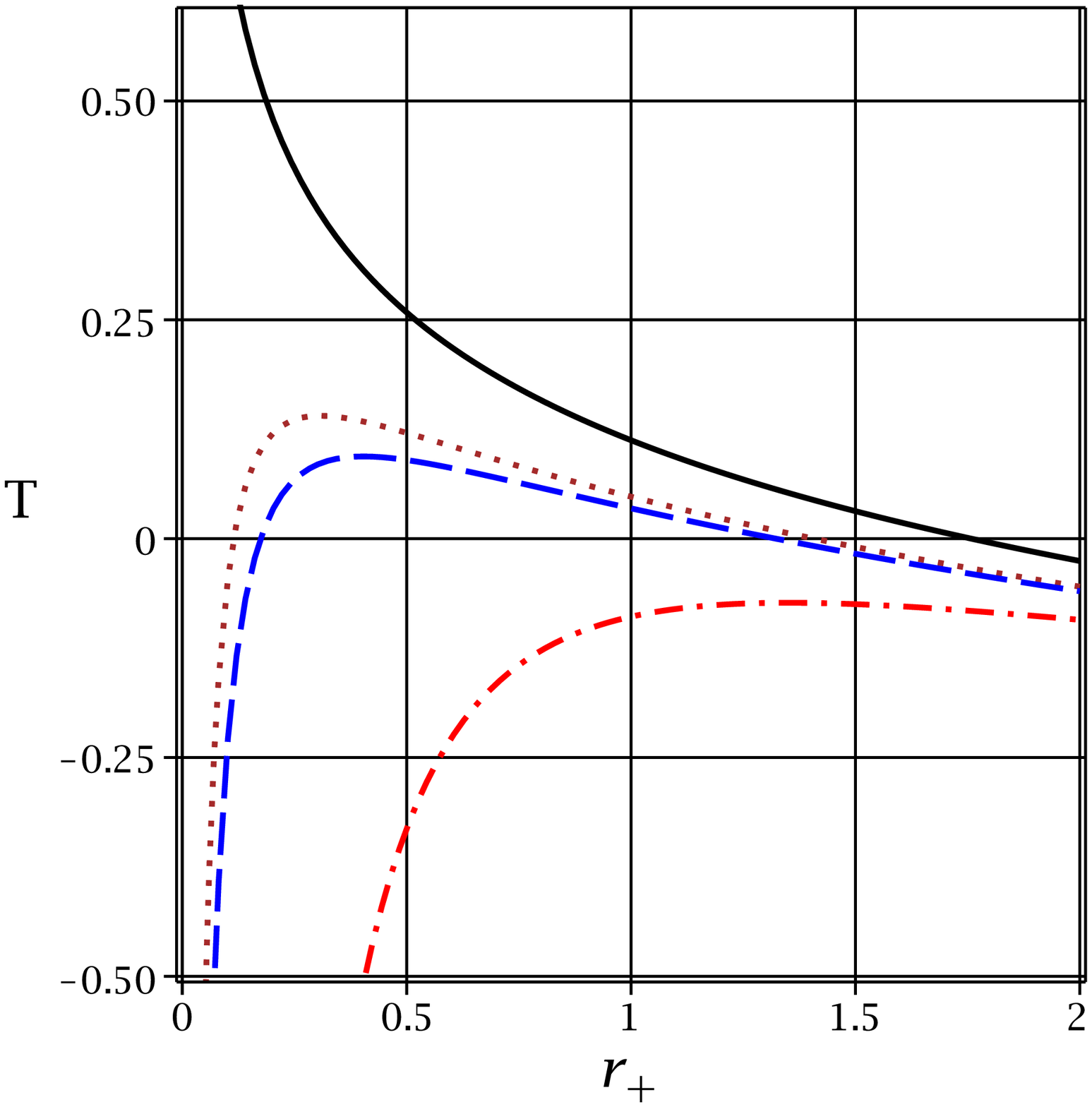} & \epsfxsize=5.5cm \epsffile{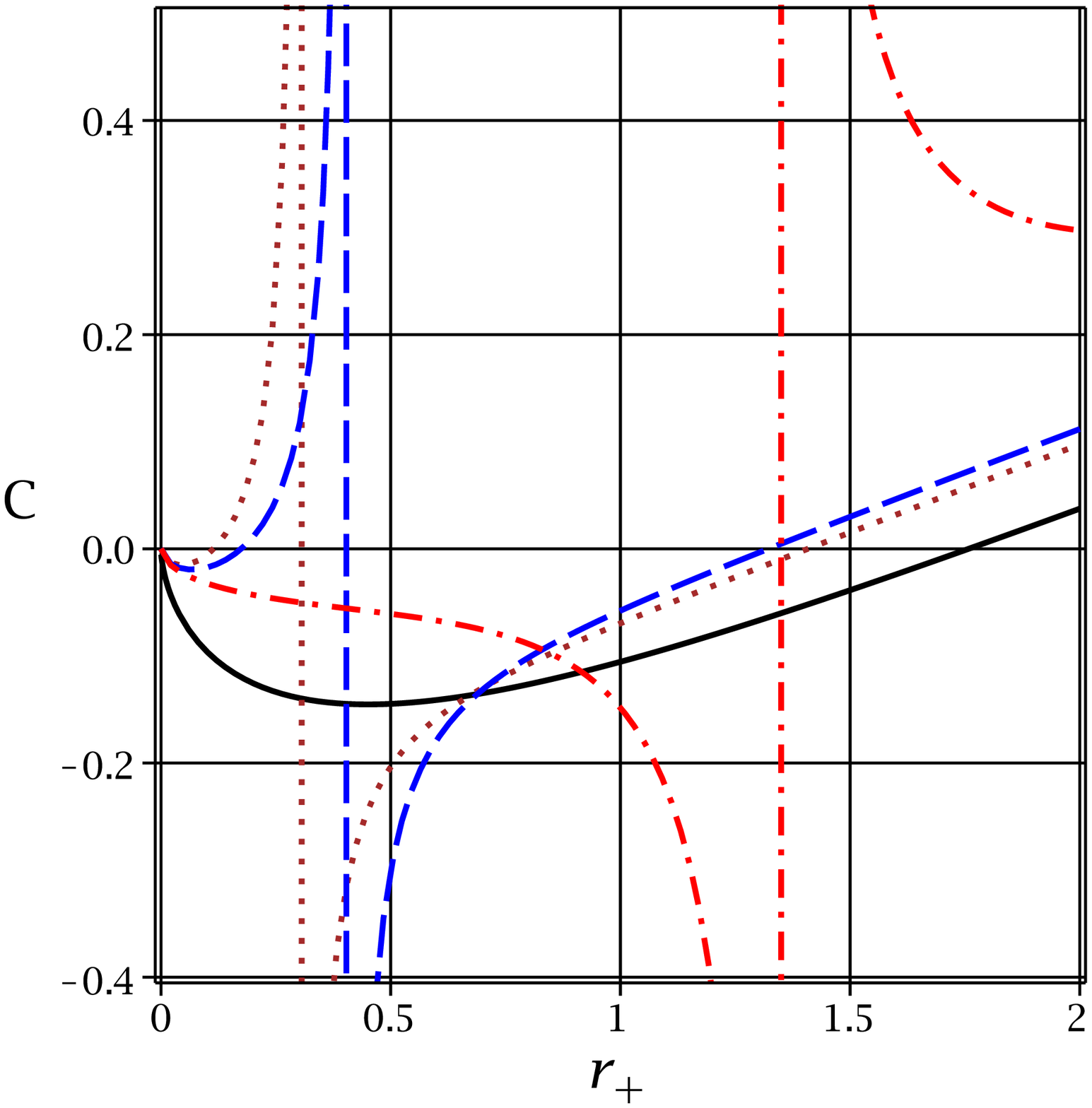}%
\end{array}
$%
\caption{$T$ (left panels) and $C$ (right panels) versus $r_{+}$ for $k=1$, $%
m=5$, $\Lambda=-1$, $b=1$, $q=1$, $f(\protect\varepsilon)=g(\protect%
\varepsilon)=0.9$, $\protect\beta=0$ (continuous line), $\protect\beta=0.15$
(dotted line), $\protect\beta=0.1898$ (dashed line) and $\protect\beta=2$.
\newline
Up panels: $\protect\alpha=0.5$; down panel: $\protect\alpha=1.5$.}
\label{Fig6}
\end{figure*}

\begin{figure*}[tbp]
$%
\begin{array}{cc}
\epsfxsize=5.5cm \epsffile{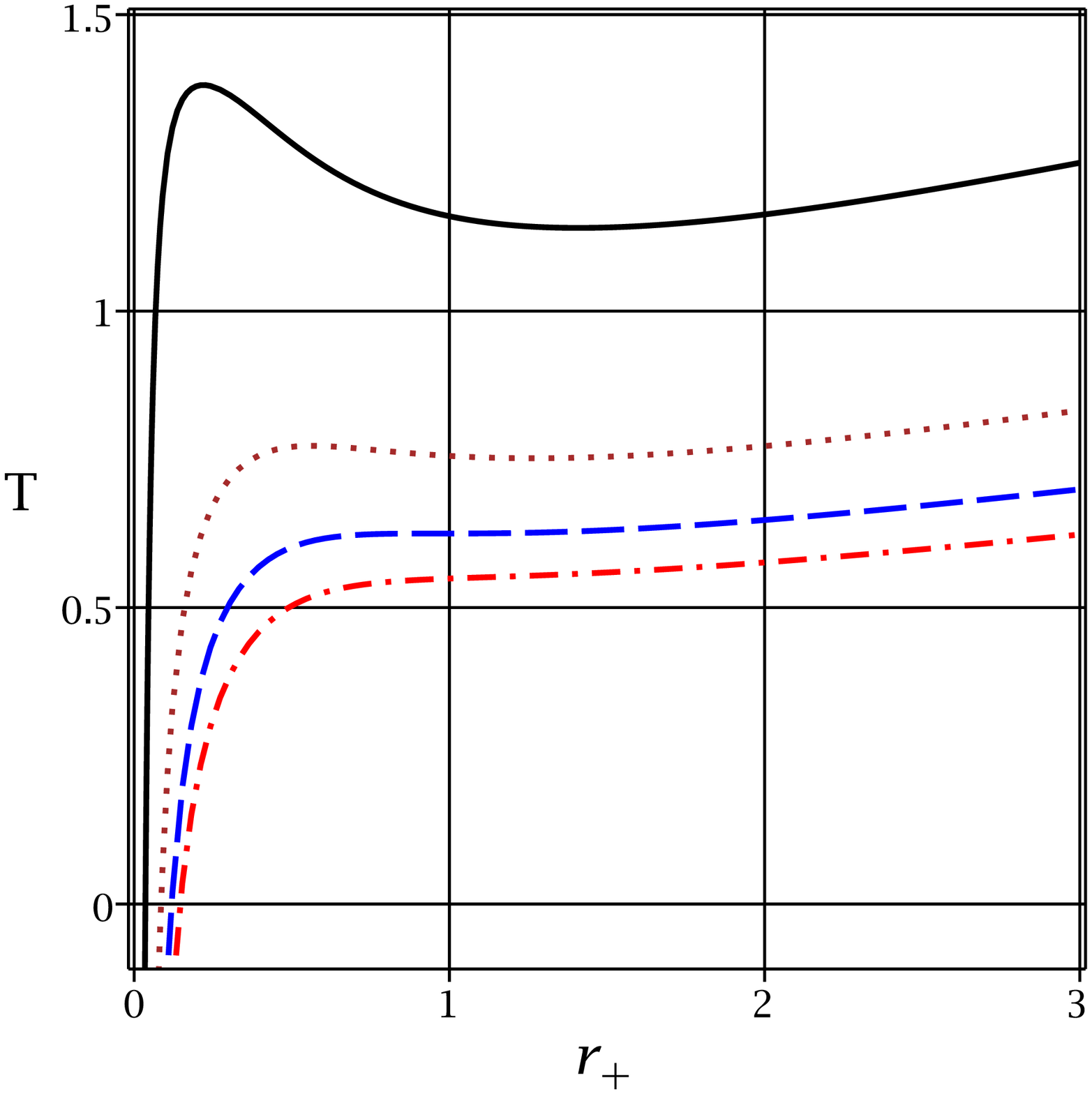} & \epsfxsize=5.5cm \epsffile{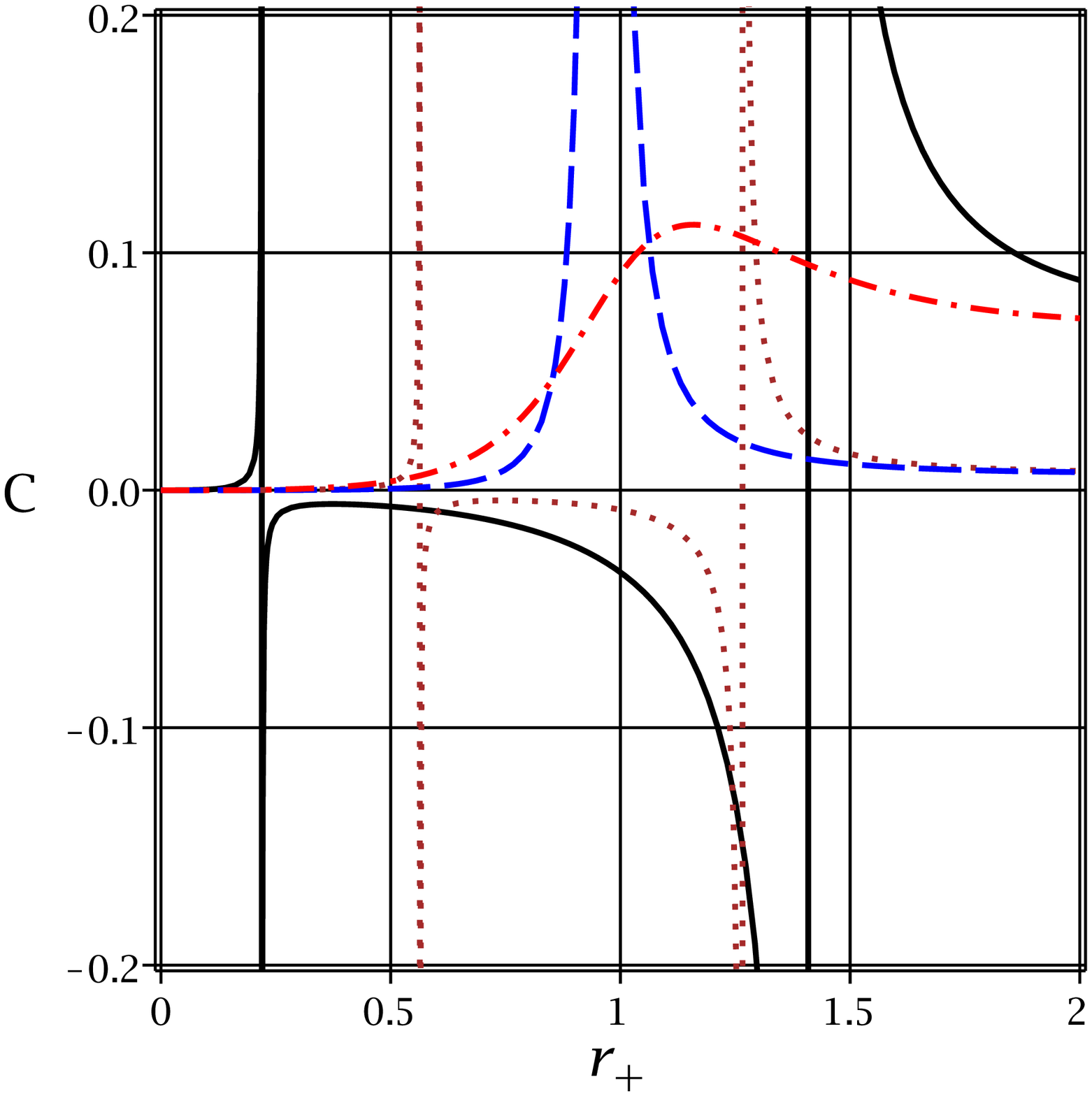}
\\
\epsfxsize=5.5cm \epsffile{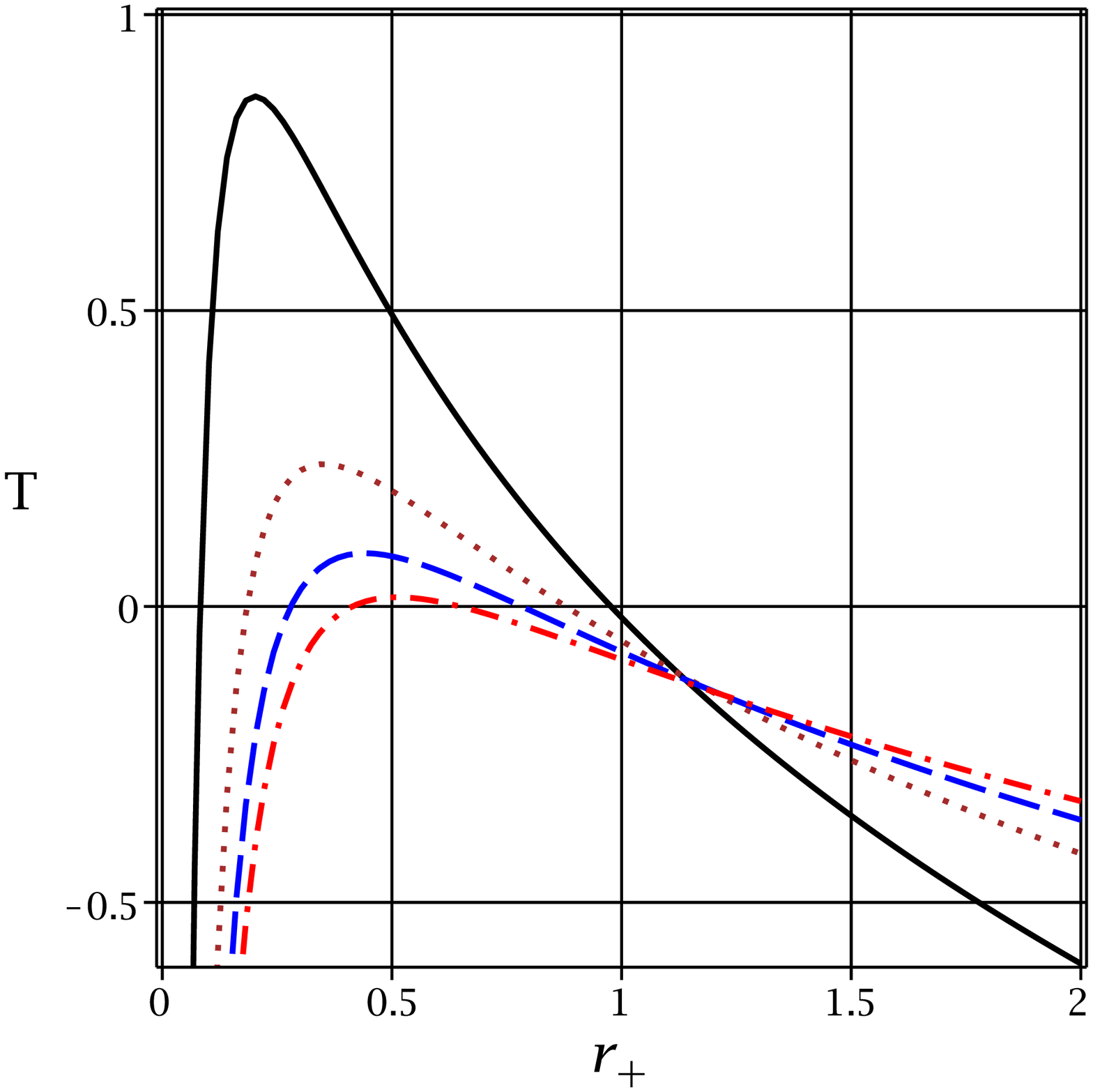} & \epsfxsize=5.5cm \epsffile{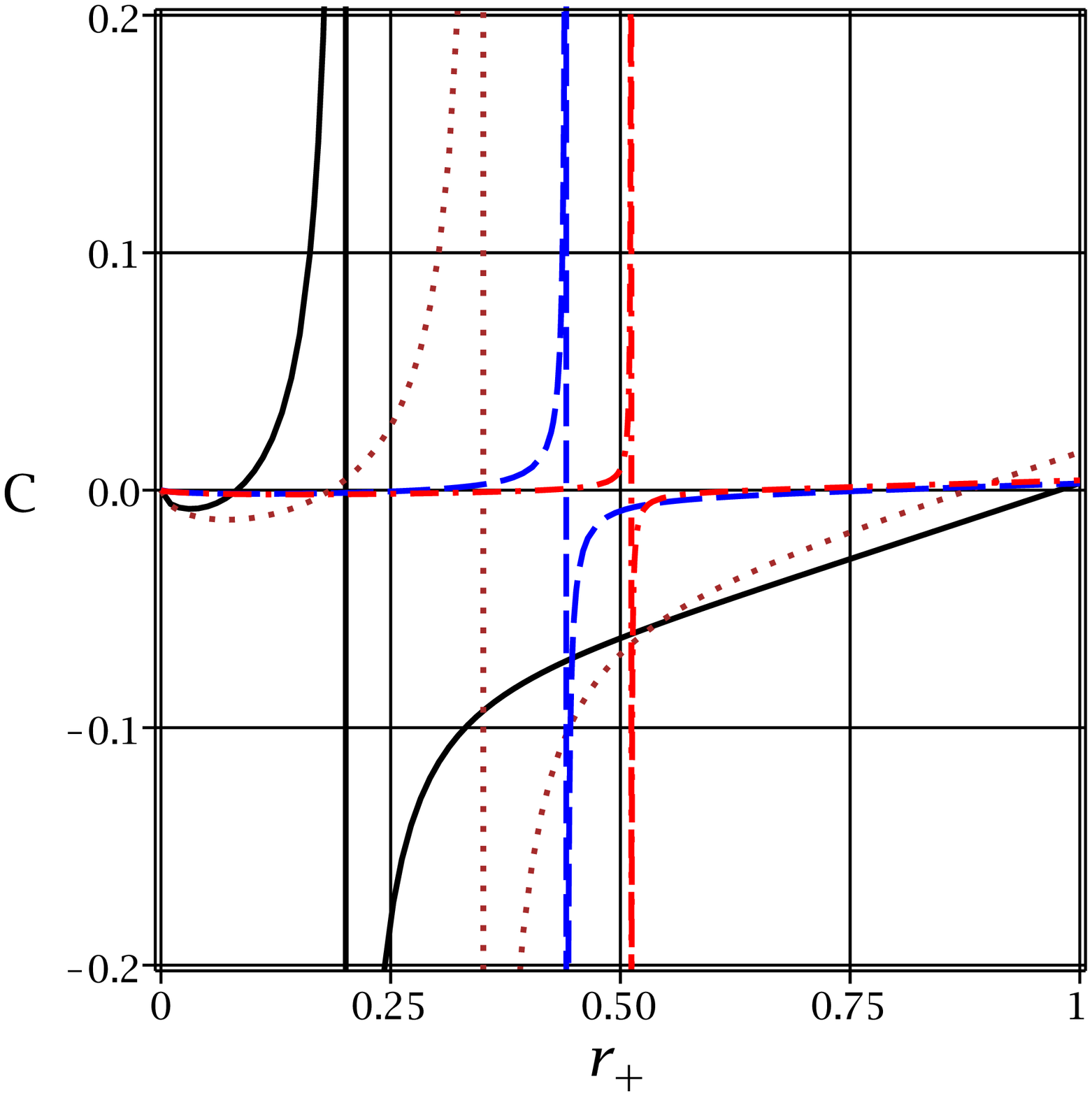}%
\end{array}
$%
\caption{$T$ (left panels) and $C$ (right panels) versus $r_{+}$ for $k=1$, $%
m=5$, $\Lambda=-1$, $b=1$, $q=1$, $\protect\beta=1$, $g(\protect\varepsilon%
)=1.1$, $f(\protect\varepsilon)=0.2$ (continuous line), $f(\protect%
\varepsilon)=0.3$ (dotted line), $f(\protect\varepsilon)=0.3567$ (dashed
line) and $f(\protect\varepsilon)=0.4$. \newline
Up panels: $\protect\alpha=0.5$; down panel: $\protect\alpha=1.5$.}
\label{Fig7}
\end{figure*}


In brief, we should note that for $\alpha <1$, we had three cases for the
heat capacity: I) existence of only one root ($0<\alpha <\alpha _{c}$); for
this case, the variation of correction parameter leads to modification in
place of the root which is an increasing function of $\zeta $ (up-left panel
of Fig. \ref{Fig4}), II) existence of one root and one divergency ($\alpha
=\alpha _{c}$); for small values of the first order correction parameter, $%
\zeta $, the place of root is modified to larger values, but the sign of
heat capacity around divergence point remains positive. Interestingly, for
sufficiently large values of the correction parameter, the place of root
will be shifted to after the divergency and the sign of heat capacity before
it and also around divergence point will be negative (up-middle panel of
Fig. \ref{Fig4}). Therefore, the only thermally stable phase is after the
root and around the divergence point, phase transition is between two
unstable black holes, III) Existence of one root and two divergencies ($%
\alpha _{c}<\alpha $); in this case, for small values of the correction
parameter, the place of root is modified to larger values of the horizon
radius but before the divergencies. The stability conditions remain
unchanged. But for medium values of the correction parameter, the root will
be located between the divergencies (up-right panel of Fig. \ref{Fig4}).
Interestingly, before smaller divergency, and between the root and larger
divergency, the heat capacity is negative. Whereas, between the root and
smaller divergency, and also after the larger divergency, the heat capacity
is positive (dashed and dashed-dotted lines in up-right panel of Fig. \ref%
{Fig4}). Increasing the value of correction parameter leads to root being
place after the larger divergency. In this case, the only stable phases
(positive heat capacity) are between the divergencies and after the root
(bold continuous and bold dotted lines in up-right panel of Fig. \ref{Fig4}).

For $\alpha >1$, we only consider the physical case in which the temperature
has two roots and between them, temperature has positive values. Evidently,
contribution of the first order correction results into modification of the
place of root for the heat capacity to higher values of the horizon radius.
The root will be located after divergency (up panels of Fig. \ref{Fig5}).
Before root (around divergence point as well) the heat capacity is negative.
The positive heat capacity could be observed after its root. But the root of
heat capacity is located after larger root of the temperature, therefore,
even though the heat capacity changes its sign to positive after its root,
but it is within non-physical region. This results into a conclusion that
existence of the first order correction for this case leads to instability
of the black holes within physical region. We see that contribution of the
first order correction destabilizes the solutions.

Our next study here is measuring the modifications in entropy by studying
the fluctuation in temperature. To do so, we have plotted diagrams for both
cases of $\alpha<1$ and $\alpha>1$ corresponding to those plotted for
temperature and heat capacity. In addition, we have plotted other diagrams
to understand the effects of first order correction on thermodynamical
behavior of the entropy as a function of temperature.

For $\alpha <1$ which is interpreted as AdS case, evidently, three
distinctive behaviors could be observed for the entropy. These behaviors
could be divided and addressed by $\alpha _{c}$ which was introduced before
in the context of heat capacity. For $0<\alpha <\alpha _{c}$, the general
behavior of the entropy is not modified on a significant level. Here,
entropy is an increasing function of temperature (continuous and dotted
lines in right panel of Fig. \ref{Fig2}). By setting $\alpha =\alpha _{c}$,
an extremum is formed for entropy versus temperature diagrams (dashed line
in right panel of Fig. \ref{Fig2}). This case corresponds to the cases where
the temperature and the heat capacity enjoy an extremum and divergency,
respectively, in their structures. Therefore, one can conclude that the
entropy of extremum point is the critical entropy where system goes under
thermal phase transition. Increasing the dilatonic parameter to reach the
region of $\alpha _{c}<\alpha $ results into formation of two extrema which
could be recognized by $T_{1}$ and $T_{2}$ (dashed-dotted line in right
panel of Fig. \ref{Fig2}). Between these two indications, the temperature is
a decreasing function of entropy and for every temperature, there exists
three different entropies. For $T=T_{1},T_{2}$ every temperature has two
specific entropies. This case corresponds to the existence of two
divergencies in the heat capacity. Thermodynamical principle informs us that
system is in favor of increasing its entropy. This indicates that for the
case where three (two) entropies are available for temperature, system moves
to the case, which has the highest entropy. This is indeed the
characteristic behavior of the phase transition. Therefore, we see that
measurement of the entropy as a function of the fluctuation of temperature
provides us specific characteristic that enables us to recognize critical
behavior.

For $\alpha >1$ which is interpreted as dS case, we see that for
non-physical case where system has the negative temperature, the entropy is
positive valued (continuous line in right panel of Fig. \ref{Fig3}).
Increasing dilatonic parameter leads to temperature acquires an extreme root
where, interestingly, entropy has positive and non-zero value (dotted line
in right panel of Fig. \ref{Fig3}). Increasing the dilatonic parameter
furthermore leads to formation of a region of positive temperature with a
maximum provided for temperature (dashed and dashed-dotted lines in right
panel of Fig. \ref{Fig3}). In this case, except for the maximum, every
temperature has two different entropies. The maximum of temperature is where
the system acquires divergency in its heat capacity. Now remembering that
system desires higher values of the entropy, we can see that characteristic
phase transition behavior is observed by system jumping from smaller entropy
to larger one at the same temperature. Interestingly, even for vanishing
temperature, there are two values provided for the entropy of system.

Now, let us focus on the effects of first order correction. As it was
pointed out, the case $\alpha<1$ (AdS case) enjoys larger number of the
possibilities in its structure depending on the value of dilatonic
parameter. To have more conclusive discussion regarding the results of first
order correction, we study the effects of variation of the correction
parameter for different cases provided $\alpha<1$ and for the physical case
observed in $\alpha>1$ separately; I) for $0<\alpha<\alpha_{c}$ in case of $%
\alpha<1$, we see that adding first order correction results into formation
of a minimum for entropy, $S_{min}$ (down-left panel of Fig. \ref{Fig4}). In
addition, for vanishing temperature, entropy is positive and non-zero
valued, $S_{0}$. In the region $S_{min}<S<S_{0}$, for every entropy, there
exists two different temperatures. The minimum and $S_{0}$ are increasing
function of the correction parameter, II) for $\alpha=\alpha_{c}$ in case of
$\alpha<1$, where an extremum was observed in the absence of correction
term, adding the correction results into formation of a minimum for the
entropy alongside the extremum that was observed before (down-middle panel
of Fig. \ref{Fig4}). But interestingly, the entropies of minimum and
extremum points are increasing functions of the correction parameter. Here
too, similar to previous case, for specific region of entropy, every entropy
has two different temperatures. Now, remembering that extremum point is
where system goes under phase transition, one can conclude that entropy of
the critical point depends on value of the correction parameter. This shows
that critical behavior in the presence of first order correction is reached
in higher values of the disorder in system, hence entropy, III) for $%
\alpha_{c}<\alpha$ in case of $\alpha<1$, as it was pointed out, two
distinctive temperatures where available, $T_{1}$ and $T_{2}$, in which
between them, every temperature has three different entropies. In addition,
the entropy related to $T_{1}$ is larger than entropy corresponding to $T_{2}
$ (continuous line in down-right panel of Fig. \ref{Fig4}). Now, the effects
of first order correction in this case could be divided into four groups
which are recognized with $\zeta_{1}$, $\zeta_{2}$ and $\zeta_{3}$ in which $%
\zeta_{1}<\zeta_{2}<\zeta_{3}$. For $0<\zeta<\zeta_{1}$, the general
behavior of the entropy versus temperature is same as that observed for
vanishing $T$ with one difference that entropies corresponding to $T_{1}$
and $T_{2}$ are higher than vanishing temperature (dotted line in down-right
panel of Fig. \ref{Fig4}). Increasing the correction parameter to reach the
region of $\zeta_{1}<\zeta<\zeta_{2}$, interestingly, results into formation
of cycle for entropy versus temperature (dashed line in down-right panel of
Fig. \ref{Fig4}). This means that between $T_{1}$ and $T_{2}$, there is yet
another temperature, $T^{\prime }$ that has two different entropies
corresponding to it. In this case, the entropy of $T_{1}$ is still bigger
than entropy of $T_{2}$. If we choose the correction parameter from the
region of $\zeta_{2}<\zeta<\zeta_{3}$, the cyclic behavior could be
observed, but interestingly, the entropy corresponding to $T_{2} $ is larger
than the one related to $T_{1}$ (dashed-dotted line in down-right panel of
Fig. \ref{Fig4}). This means that although the entropies related to these
two temperatures are increasing functions of the correction parameter, but
the entropy related to $T_{2}$ grows faster comparing to the one
corresponding to $T_{1}$. Finally, for the region of $\zeta_{3}<\zeta$, the
cyclic behavior is vanished, but since the entropy of $T_{2}$ is now bigger
than the one related to $T_{1}$, the diagrams has opposite curve comparing
to small values of the correction parameter (bold continuous and dotted
lines in down-right panel of Fig. \ref{Fig4}), IV) finally, for the physical
case observed in $\alpha>1$, in the absence of correction, we observed a
maximum for temperature. Interestingly, the existence of first order
correction results into a cyclic (closed) diagram for the entropy versus
temperature (down panels of Fig. \ref{Fig5}). In addition, there is a
minimum obtained for the entropy which has maximum temperature as its
correspondence. In this case, for every temperature (entropy) there exists
two entropies (temperatures). The minimum of entropy is an increasing
function of the correction parameter.

In order to complete our study, we finally investigate the effects of
nonlinearity parameter and energy functions on thermodynamical behavior of
these black holes for two branches of $\alpha<1$ (AdS case) and $\alpha>1$
(dS case).

First, we turn our attention to the nonlinear nature of solutions. For case
of $\alpha<1$, evidently, the effects of nonlinear electrodynamics could be
divided into three categories with a critical value for nonlinearity
parameter, $\beta_{c}$ (in which $c$ stands for critical). In the absence of
nonlinearity parameter, temperature has a minimum and it is positive valued
everywhere (continuous line in up-left panel of Fig. \ref{Fig6}). In the
presence of nonlinearity parameter and for the region of $\beta<\beta_{c} $,
temperature enjoys a root, a maximum and a minimum in its structure (dotted
line in up-left panel of Fig. \ref{Fig6}). As for $\beta=\beta_{c}$, the
number of extremum is reduced to one which is located after the temperature
(dashed line in up-left panel of Fig. \ref{Fig6}). Increasing the
nonlinearity parameter beyond $\beta_{c}<\beta$ results into vanishing the
extremum in temperature, and temperature will be an increasing function of
the horizon radius with a root (dashed-dotted line in up-left panel of Fig. %
\ref{Fig6}). The corresponding heat capacity diagram shows that temperature
and heat capacity share same roots and the extrema of temperature are
matched with divergencies in the heat capacity (compare up-left and up-right
panels of Fig. \ref{Fig6}). Considering this fact, one can conclude two
important points for $\alpha<1$ branch; I) for small values of the
nonlinearity parameter, thermodynamical structure of the black holes enjoys
the existence of phase transition, II) increasing the nonlinearity parameter
results into absence of the phase transition for these black holes. It is
worthwhile to mention that for vanishing the nonlinearity parameter, since
temperature has a minimum, heat capacity has a divergency, which signals for
possible critical behavior. The sign of heat capacity changes from negative
to positive around divergency which indicates a phase transition between
small unstable black holes to large stable ones. For case of $\alpha>1$, in
the absence of the nonlinearity parameter, temperature is a decreasing
function of the horizon radius with a root (continuous line in down-left
panel of Fig. \ref{Fig6}). In the presence of nonlinearity parameter,
temperature acquires a maximum with two roots. The maximum and number of the
roots are decreasing functions of the nonlinearity parameter. This indicates
that for sufficiently large values of the nonlinearity parameter, the
maximum will be relocated to negative values and roots will be vanished
(dashed-dotted line in down-left panel of Fig. \ref{Fig6}). Therefore, one
can conclude that for $\alpha>1$ branch, the effects of increasing
nonlinearity parameter result into elimination of physical solutions. As for
the stability, interestingly, in the absence of nonlinearity parameter and
in the region where temperature is positive, the heat capacity is negative,
hence solutions are thermally unstable. After the root, although heat
capacity is positive, temperature is negative valued. Therefore, for this
case, only unstable solutions exist. In the presence of nonlinearity
parameter, if the maximum of the temperature is located at positive values
(leads to presence of two roots for temperature), heat capacity enjoys a
phase transition between large unstable to small stable black holes (dotted
and dashed lines in down-right panel of Fig. \ref{Fig6}).

The situation for the effects of the energy functions is investigated in
Fig. \ref{Fig7}. Here, we have investigated the effects of gravity's rainbow
for both branches of $\alpha<1$ (up panels of Fig. \ref{Fig7}) and $\alpha>1$
(down panels of Fig. \ref{Fig7}). Evidently, for small values of the rainbow
functions, temperature enjoys a root and a maximum and a minimum in its
structure which correspondingly, heat capacity would have the same and two
divergencies matching extrema points in temperature (continuous line in up
panel of Fig. \ref{Fig7}). This shows that these black holes, for small
values of the energy functions enjoy a phase transition over a region which
is located between two divergencies of heat capacity. The number of the
extremum in temperature, hence divergency in heat capacity is a decreasing
function of the energy function. For certain value of the energy function,
temperature will have an extremum and heat capacity enjoys a divergency
(dashed lines in up panel of Fig. \ref{Fig7}). In this case, a phase
transition between two stable black holes takes place. Increasing the energy
function beyond this value results into the absence of extremum in
temperature, hence divergency in the heat capacity. Therefore, existence of
the critical behavior depends on values of the energy functions for $\alpha<1
$ branch. As for $\alpha>1$, it is evident that maximum exists for
temperature which corresponds to a divergency in heat capacity (down panels
of Fig. \ref{Fig7}). For small values of the energy function, this maximum
is within positive valued region and temperature also enjoys existence of
two roots. The maximum is a decreasing function of $f(\varepsilon)$ and for
sufficiently large values of this energy function, it will be relocated into
negative region indicating negative temperature without root, hence absence
of physical solutions. Therefore, one can conclude that existence of
physical solutions for $\alpha>1$, depends on the values that energy
functions can acquire.

\section{Conclusion}

Regarding the fact that we should consider high energy (UV) regime near the
black holes motivates us to consider an energy dependent spacetime with a
minimal coupling between gravity, dilaton scalar field as well as a
nonlinear $U(1)$ gauge field. In this paper, we have studied the black hole
solutions in dilaton gravity's rainbow in the presence of BI source. It was
shown that the nonlinearity parameter, energy functions and dilatonic
parameter modify the type of the singularity, the places and number of the
horizon radius and their type as well. These modifications highlighted the
importance of matter fields and gravities.

Next, we obtained conserved and thermodynamical quantities and we proved
that despite the mortifications of the gravity's rainbow, dilatonic gravity
and BI field, the first law of the black holes thermodynamics is valid.

Our study in the context of temperature and heat capacity provided us with
interesting properties for dilatonic parameter. First of all, we were able
to impose an specific limit on dilatonic parameter in order to avoid
divergent temperature. In addition, it was shown that for $\alpha <1$, the
characteristic behaviors of temperature and heat capacity are exactly the
same as those that were observed for black holes in AdS spacetime. Whereas
for $\alpha >1$, the properties extracted for the temperature and the heat
capacity matches those observed for black holes in dS case. These two
specifications provide us with the possibility of conducting studies that
are specified for the branch AdS/dS such as AdS/CFT correspondence.

We have investigated the effects of thermal fluctuations on thermodynamical
quantities. Although the entropy and mass were affected by thermal
fluctuation, the first law remained valid in this case as well. In studying
the effects of first order correction, it was shown that although the phase
transition points are independent of first order correction, the stability
conditions, which determine thermal structure of the black holes, are highly
sensitive to variation of the first order correction parameter. Here, we
observed that depending on the choice for correction parameter, the
stability regions for black holes are modified on high level depending on
the possible scenarios provided for $\alpha <1$. On the other hand, the
contribution of first order correction resulted into destabilization of the
solutions for $\alpha >1$ case. Remembering that $\alpha <1$ could be
interpreted as AdS spacetime and $\alpha >1$ as dS case, one can conclude
that for the AdS case, the stability conditions depend on the correction
parameter while for dS case, first order correction makes the solutions
unstable within physical region. Furthermore, one can observe that
generalization to include the first order correction, results into larger
families of thermal structures provided for these black holes in AdS case.

The measurement of entropy as a function of the fluctuation in temperature
revealed profound deviation from the non-correction case. In $\alpha<1$, it
was shown that depending on choices of the correction parameter, it was
possible to introduce phenomena such as cyclic behavior and transition from
one specific curve for diagrams to opposite curve. As for $\alpha>1$,
existence of first order correction resulted into cyclic behavior as well
for entropy versus temperature diagrams and a minimum for entropy which has
maximum of temperature as its correspondence. Generally speaking, the
possibilities of such behaviors were provided due to the fact that entropy
is correction dependent while the temperature is independent of it. In
addition to mentioned important effects of the first order correction, there
is another effect that is of importance; the temperatures of the critical
point through entropy versus temperature diagrams remained fixed but the
critical entropies where shifted to higher values. This confirms that
generically, the black holes with first order correction included, go under
phase transition in higher level of disorder in their system.

To complete our thermodynamical investigation, we studied the effects of the
nonlinearity parameter and energy functions on temperature and heat
capacity. In $\alpha<1$ case, interestingly, it was observed that for small
values of the energy function and nonlinearity parameter, system enjoys a
phase transition over a region. Increasing these quantities leads to
formation of an extremum for temperature, hence a divergency for heat
capacity. In this case, the phase transition was taking place over a single
point. As for $\alpha>1$, the maximum for temperature was a decreasing
function of the energy functions and nonlinearity parameter, and for
sufficiency large values of these quantities, maximum will be in negative
region of temperature which is interpreted as absence of physical solutions.
It is worthwhile to mention that for vanishing nonlinearity parameter, the
general behavior of temperature and type of divergencies where completely
different indicating the high contribution of the nonlinear electromagnetic
field generalization.

Our study in this paper confirmed that consideration of the first order
correction, results into larger classes of thermodynamical structure
provided for black holes. But there are two questions that should be
answered: I) are we free to choose any value for the correction parameter?
II) are all possibilities that are provided for the thermal structure of the
black holes, physical ones? One method to regard these questions is through
the extended phase space. The concept of extended phase space provides the
possibility of studying specific properties for black holes such as
compressibility coefficient and speed of sound. These properties along with
the concept that speed of sound must not exceed the speed of light, provide
us with the possibility of finding upper/lower limit on values that
correction parameter can acquire and distinguish physical thermodynamical
structures from non-physical ones that were obtained in this paper. The only
shortcoming of this method is the fact that extended phase space could only
be introduced for AdS case. In any case, this issue is now under
investigation and we hope to address these questions in another paper.

\begin{acknowledgements}
We thank the Shiraz University Research Council. This work has
been supported financially by the Research Institute for Astronomy
and Astrophysics of Maragha, Iran.
\end{acknowledgements}

\appendix

\section{Einstein BI AdS black holes}

The metric function of $4-$dimensional black holes in the presence of
Born-Infeld electromagnetic field and cosmological constant is given by \cite%
{Hello}
\begin{eqnarray}
\Psi(r)&=&k-\frac{m}{r}+\left( \frac{2\beta ^{2}-\Lambda }{3}\right) r^{2}-
\notag \\
&&\frac{2\beta ^{2}r^{2}}{3}\sqrt{1+\frac{2q^{2}}{\beta ^{2}r^{4}} }+\frac{%
8q^{2}\mathcal{H}}{3r^{2}} ,  \label{f(r)}
\end{eqnarray}%
in which $\mathcal{H}$ is the following hypergeometric function
\begin{equation}
\mathcal{H}= {}_{2}\mathcal{F}_{1}\left( \left[ \frac{1}{2},\frac{1}{4}%
\right] ,\left[ \frac{5}{4}\right] ,-\frac{2q^{2}}{\beta ^{2}r^{4}} \right) ,
\label{H}
\end{equation}
where $q$ and $m$ are two integration constants related to the electric
charge and total mass of the black hole, respectively.

\begin{figure*}[tbp]
$%
\begin{array}{cc}
\epsfxsize=5cm \epsffile{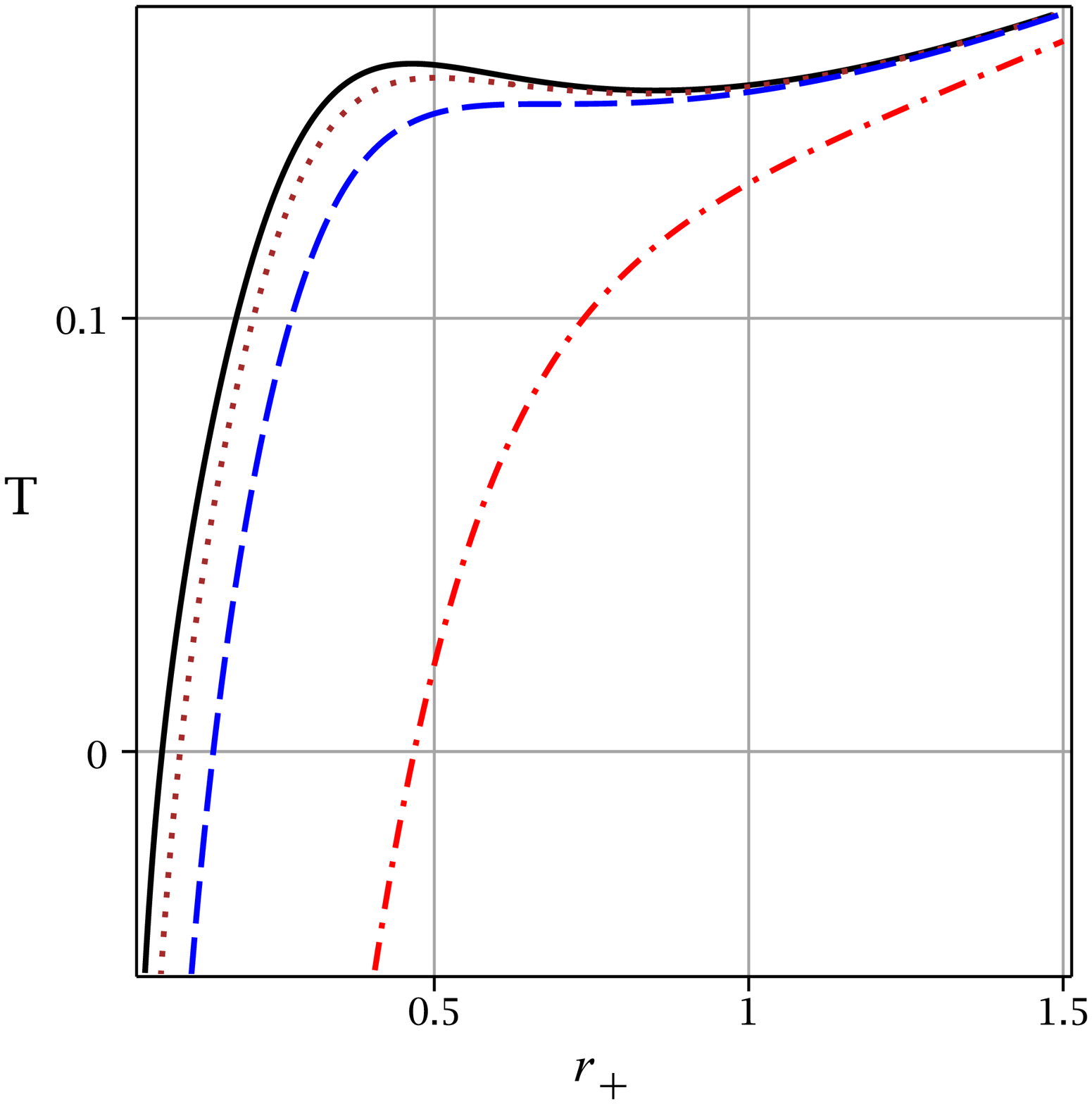} & \epsfxsize=5cm \epsffile{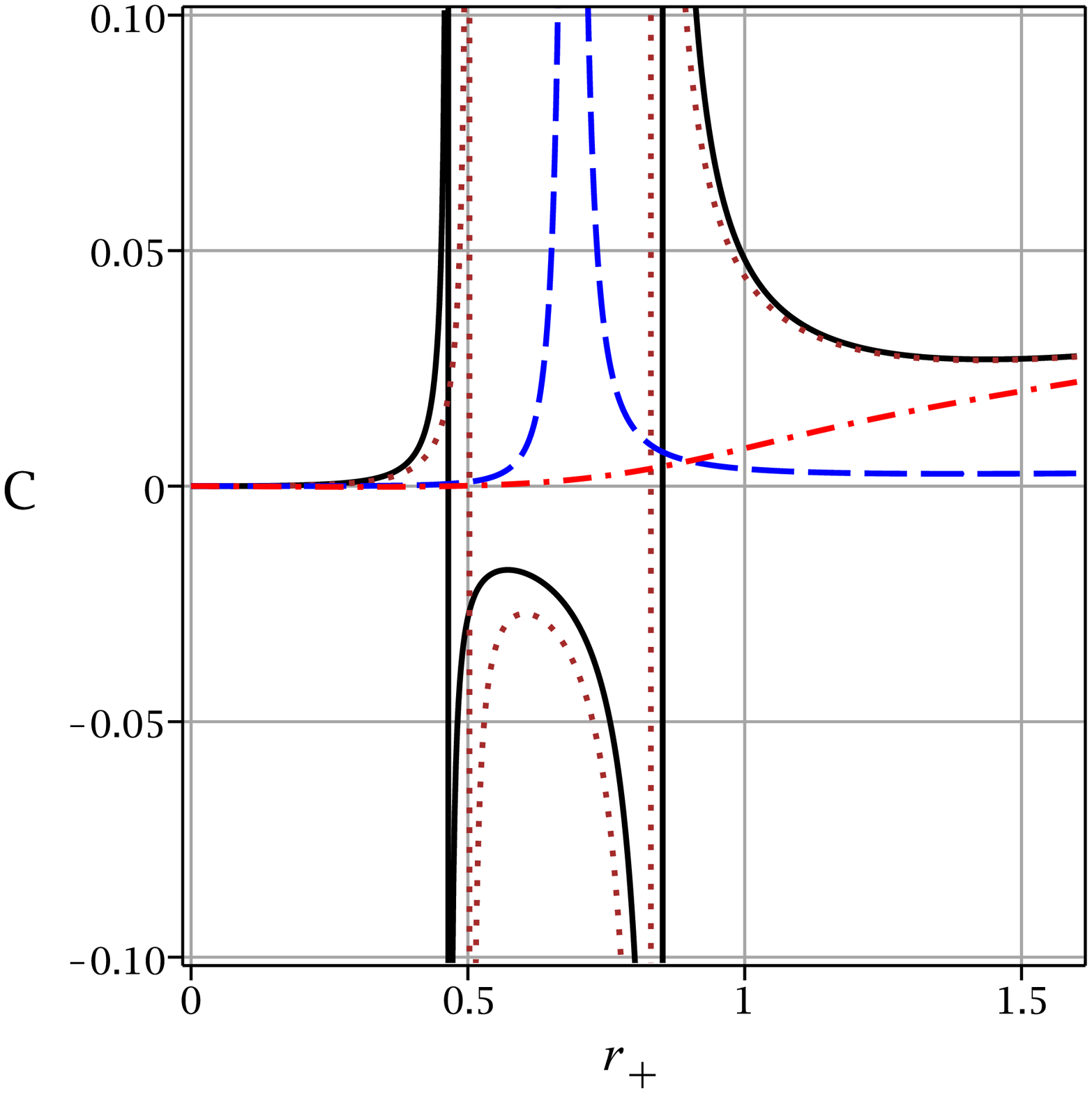}
\\
\epsfxsize=5cm \epsffile{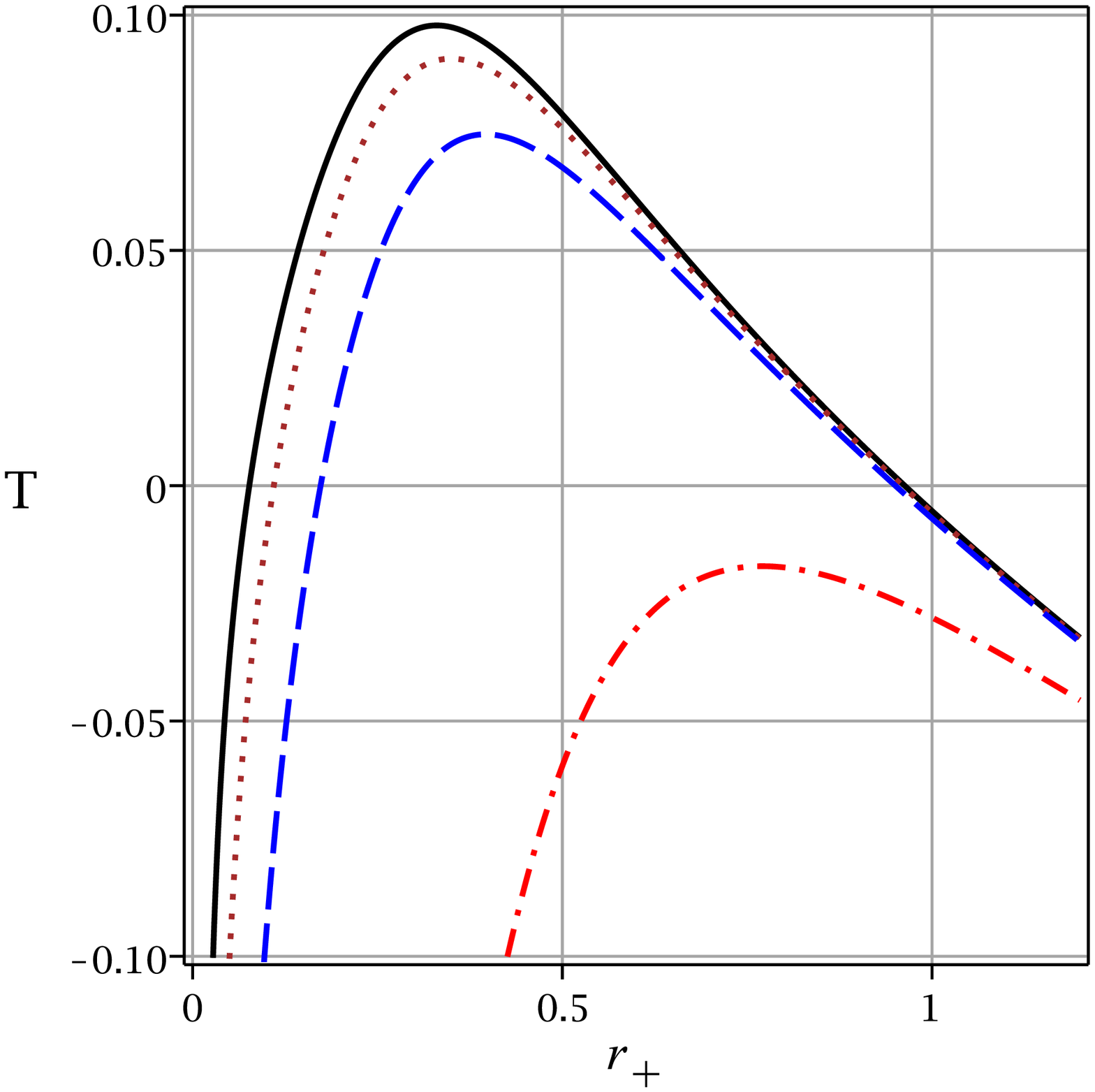} & \epsfxsize=5cm \epsffile{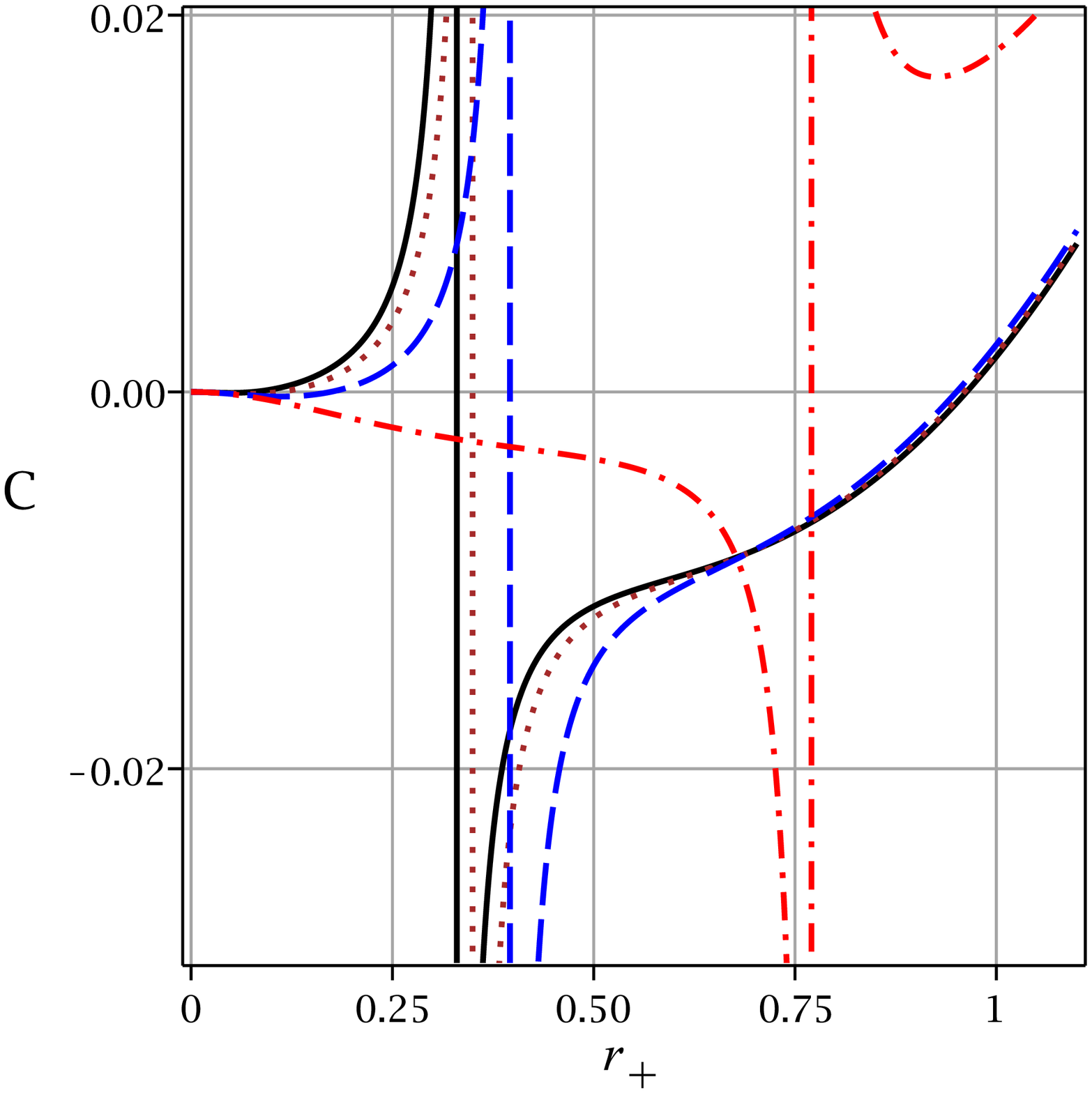}%
\end{array}
$%
\caption{$T$ (left panel) and $C$ (right panel) versus $r_{+}$ for $\protect%
\beta =2$ and $k=1$; $q=0.26$ (continues line), $q=0.27$ (dotted line), $%
q=0.2964$ (dashed line) and $q=0.6$ (dashed-dotted line). \newline
Up panels: $\Lambda=-1$; down panels: $\Lambda=1$.}
\label{FigFig1}
\end{figure*}


The temperature of these black holes could be found as \cite{Hello}
\begin{equation}
T=\frac{k}{4\pi r_{+}}+\frac{\left( 2\beta ^{2}-\Lambda \right) r_{+}}{4\pi }%
-\frac{\beta ^{2}r_{+}}{2\pi }\sqrt{1+\frac{2q^{2}}{\beta ^{2}r_{+}^{4}}} ,
\label{TotalTT}
\end{equation}%
where $\Gamma _{+}=\frac{2q^{2}}{\beta ^{2}r_{+}^{4}}$.

The entropy of these black holes are obtained as \cite{Hello}
\begin{equation}
S=\frac{1}{4}r_{+}^{2},  \label{TotalS}
\end{equation}%
and the heat capacity is given by%
\begin{equation}
C_{Q}=\frac{T}{\left( \frac{\partial ^{2}M}{\partial S^{2}}\right) _{Q}}=%
\frac{T}{{\left( \frac{\partial T}{\partial S}\right) _{Q}}}.  \label{CQ}
\end{equation}

Considering Eqs. (\ref{TotalTT}) and (\ref{TotalS}), it is a matter of
calculation to show that \cite{Hello}
\begin{eqnarray}
{\left( \frac{\partial T}{\partial S}\right) _{Q}}&=&-\frac{k}{2r_{+}^{3}}+%
\frac{2\beta ^{2}-\Lambda }{2\pi r_{+}}-\frac{\beta ^{2}}{\pi r_{+}}\left( 1+%
\frac{2q^{2}}{\beta ^{2}r_{+}^{4}}\right) ^{\frac{3}{2}}+  \notag \\
&&\frac{2q^{2}\left( 3+\frac{2q^{2}}{\beta ^{2}r_{+}^{4}}\right) }{\pi
r_{+}^{5}\sqrt{1+\frac{2q^{2}}{\beta ^{2}r_{+}^{4}}}}.
\end{eqnarray}

The thermodynamical behavior of the AdS black holes and dS ones are given in
Fig. \ref{FigFig1}.

\end{document}